\documentclass[structabstract]{aa}  
\usepackage{graphicx}
\usepackage{txfonts}
\newcommand{\pa}{\partial}
\newcommand{\ob}{\overline}
\newcommand{\beq}{B_{\rm eq}}
\newcommand{\erf}{\textrm{erf}}
\newcommand{\Rs}{R_{\odot}}
\newcommand{\Rm}{R_{\rm m}}
\begin{document}
\title{Magnetic helicity fluxes in interface and flux transport dynamos}
   \author{Piyali Chatterjee\inst{1}, Gustavo Guerrero\inst{1}
          \and
          Axel Brandenburg
          \inst{1,2}
          }

   \institute{NORDITA, AlbaNova University Center, Roslagstullsbacken 23,
              SE 10691 Stockholm, Sweden;
              \email{piyalic@nordita.org}
   \and
              Department of Astronomy, AlbaNova University Center,
              Stockholm University, SE 10691 Stockholm, Sweden
             }
   \titlerunning{Magnetic helicity fluxes in spherical shell dynamos}
   \authorrunning{P. Chatterjee et al.}

   \date{$ $Revision: 1.86 $ $}

 
  \abstract
   {
    Dynamos in the Sun and other bodies tend to produce magnetic fields
    that possess magnetic helicity of opposite sign at large and small
    scales, respectively.
    The build-up of magnetic helicity at small scales provides an important
    saturation mechanism.
   }
   {
    In order to understand the nature of the solar dynamo we need to
    understand the details of the saturation mechanism in spherical
    geometry.
    In particular, we want to understand the effects of magnetic
    helicity fluxes from turbulence and meridional circulation.
    }
   {
    We consider a model with just radial shear confined to a thin layer
    (tachocline) at the bottom of the convection zone.
    The kinetic $\alpha$ owing to helical turbulence is assumed to be localized
    in a region above the convection zone.
    The dynamical quenching formalism is used to describe the build-up
    of mean magnetic helicity in the model, which results in a magnetic
    $\alpha$ effect that feeds back on the kinetic $\alpha$ effect.
    In some cases we compare with results obtained using a simple
    algebraic $\alpha$ quenching formula.
    }
   {
   In agreement with earlier findings, the magnetic $\alpha$ effect
   in the dynamical $\alpha$ quenching formalism has the opposite sign
   compared with the kinetic $\alpha$ effect and leads to a catastrophic
   decrease of the saturation field strength with increasing magnetic
   Reynolds numbers.
   However, at high latitudes this quenching effect can lead to secondary
   dynamo waves that propagate poleward due to the opposite sign of $\alpha$.
   Magnetic helicity fluxes both from turbulent mixing and from
   meridional circulation alleviate catastrophic quenching.
   }
   {}

\maketitle
%

\section{Introduction}
The solar dynamo models developed so far and which agree with solar
magnetic field observations tend to solve the $\alpha\Omega$ mean
field dynamo equations. 
The turbulent $\alpha$-effect first proposed by  Parker (1955) is
believed to be generated due to helical turbulence in the convection
zone of the Sun.  Since $\alpha$ is generated due to quadratic
correlations of the 
small-scale turbulence we need a closure in order to complete
the set of mean field equations, e.g., the first order 
smoothing approximation (FOSA), and express the mean electromotive 
force in terms of the mean magnetic fields. This turbulent $\alpha$
encounters a critical problem when the energy of the mean field
becomes comparable to the equipartition energy of the turbulence in
the convection  zone and hence it becomes increasingly difficult for
the helical turbulence to twist rising blobs of magnetic field. The
solar dynamo modellers have traditionally used  what is referred to as
algebraic alpha quenching to mimic this phenomena.  This involves
replacing $\alpha$ by $\alpha_0/(1+\ob{B}^2/B_{\rm eq}^2)$,  an
expression used since Jepps (1975), or by $\alpha_0/(1+R_{\rm m}
\ob{B}^2/B_{\rm eq}^2)$, where $\alpha_0$ is the unquenched value and
$R_{\rm m}$ is the magnetic  Reynolds number, $\ob{B}$ is the mean
magnetic field and $B_{\rm eq}$ is the equipartition magnetic field. 
The latter expression has been discussed since the early work of
Vainshtein \& Cattaneo (1992). The $R_{\rm m}$ in the denominator
comes from  the fact that the  small-scale fluctuating magnetic field
reaches equipartition long  before the mean magnetic field does.
This has been supported by several numerical  experiments to determine
the saturation behaviour of $\alpha$ (e.g. Cattaneo \& Hughes 1996,
Ossendrijver et al. 2002). Given the large magnetic Reynolds numbers
of Astronomical objects, such phenomena is referred to as
catastrophic quenching. 
 
After the discovery of the layer of strong radial shear (called the
tachocline by Spiegel \& Zahn 1992) at the bottom of the solar 
convection zone, Parker (1993) proposed a new class of solar dynamo
models called the interface dynamo. In these models the shear is
confined to a narrow overshoot layer just beneath the convection zone,
also the region of $\alpha$ effect. The dynamo wave propagates in a
direction given by the Parker--Yoshimura rule at the interface between
the two layers defined by a steep gradient in the turbulent
diffusivity. The toroidal field produced due to stretching by the
shear is much stronger than the poloidal field and remains confined in
the overshoot layer, away from the region where the $\alpha$ effect
operates.  It may be noted that the interface dynamo
model may have serious problems when solar-like rotation with positive
latitudinal shear is included (Markiel \& Thomas 1999). 
Similarly, in the Babcock-Leighton class of flux transport models
(Choudhuri et al.\ 1995; Durney 1995) the toroidal and the poloidal
fields  are produced in two different layers. Unlike in the interface
dynamo models, the coupling between the two layers is mediated both by
diffusion and the conveyer belt mechanism of the meridional
circulation. 

It has been proposed that in interface and Babcock-Leighton type
dynamos,  the $\alpha$ effect is not catastrophically quenched  
at high $\Rm$ because the strength of the toroidal field is 
very weak in the region of finite turbulent $\alpha$ 
(e.g. Tobias, 1996; Charbonneau, 2005).  
However, according to our knowledge, not
much has been done to study the variation of the amplitude of the
saturation magnetic field with the magnetic Reynolds  number for
these classes of  $\alpha\Omega$ dynamos.  Zhang et al (2006) made an
attempt to reproduce the surface  observations of current helicity in
the Sun using a 2D mean field  dynamo model in spherical coordinates
coupled with the dynamical  quenching equation. In a separate paper
(Chatterjee, Brandenburg \&  Guerrero, 2010) we have demonstrated that
interface dynamo models are  also subject  to catastrophic quenching. 

It has been identified a decade ago that the small-scale
magnetic helicity generated due to the dynamo action back reacts on
the helical turbulence and quenches the dynamo (Blackman \& Field,
2000; Kleeorin et al. 2000). It has now been shown that this
mechanism reduces the saturation
amplitude of the magnetic field ($B_{sat}$) with increasing
magnetic Reynolds number ($\Rm$). Nevertheless this constraint 
may be lifted if the system is able to get rid of small scale
helicity through several ways like open boundaries, advective, 
diffusive and shear driven fluxes 
(Shukurov et al. 2006, Zhang et al. 2006, Sur et
al. 2007, K\"apyl\"a et al. 2008, Brandenburg et al. 2009, Guerrero et
al. 2010).    
Even though the helicity constraint in direct numerical
simulations (DNS) of dynamos with strong shear have been clearly
identified, the results can be matched with mean field models having a
weaker algebraic quenching than $\alpha^2$ dynamos (Brandenburg et
al.\ 2001). It is possible to include this process in mean-field 
dynamo models through an equation describing the evolution of 
the small scale current helicity. We shall refer to this equation as the    
dynamical quenching mechanism. 

In this paper we perform a series of calculations with mean 
field $\alpha\Omega$ models in spherical geometry along with a
dynamical equation for the evolution of $\alpha$ for magnetic Reynolds
numbers in the range $1\le R_{\rm m}\le 2\times10^5$. 
An important feature of the calculation is that the region of strong
narrow shear is separated from the region of helical turbulence.  
This paper in addition to providing detailed results not mentioned in
Chatterjee, Brandenburg \& Guerrero (2010), is also aimed at studying
somewhat more complicated models including meridional circulation.   
The role of diffusive helicity fluxes modelled into the dynamical
quenching equation by using a Fickian diffusion term is also discussed
for various models. It may be mentioned that helicity fluxes across  
an equator can indeed be modelled by such a diffusion term as shown by
Mitra et al. (2010). In \S2 we discuss the features of the
$\alpha\Omega$ model used, and the formulation of dynamical $\alpha$
quenching. The results are highlighted in \S3 and  conclusions are
drawn in \S4. 


\section{The basic $\alpha\Omega$ Dynamo Model} 
\subsection{Simple two-layer dynamo}
We solve the induction equation in a spherical shell assuming
axisymmetry. Our dynamo 
equations consists of the induction equations for the mean poloidal
potential $A_{\phi}(r, \theta)$ and the mean toroidal field 
$B_{\phi}(r, \theta)$. Axisymmetry demands that for all variables 
$\pa/\pa \phi = 0$. 
Let us first do a qualitative estimate of the turbulent $\alpha$ and
the turbulent diffusivity $\eta_{\rm t}$. From mixing length theory we have
(cf.\ Sur et al.\ 2008),
\begin{displaymath}
\eta_{\rm t} = \frac{u_{\rm rms}}{3k_{\rm f}},
\end{displaymath}
where $u_{\rm rms}$ is the rms velocity of the turbulent eddies, 
$k_{\rm f}$ is the wavenumber of the energy-carrying eddies, corresponding
to the inverse pressure
scale height near the base of the convection zone. 
Since we have made use of the error function profile extensively, 
let us denote 
$$\Theta^{\pm} (r,r_c,d_c)=1\pm\erf\left(\frac{r-r_{e}}{d_e}\right).$$
We have used a smoothed step profile for $\eta_{\rm t}$ given by
\begin{equation}
\label{eq:eta}
\eta(r)=\eta_r+\eta_{\rm t}\Theta^+(r,r_e,d_e)
\end{equation}
where $r_{e}=0.73\Rs$, and $d_e=0.025\Rs$. In this paper we define 
the magnetic Reynolds number $R_{\rm m}=\eta_{\rm t}/\eta_r$.
Using FOSA we also have $\alpha_0=\tau \epsilon_{\rm f} \omega_{\rm
  rms}u_{\rm rms}/3$,  
where $\omega_{\rm rms}$ is the rms vorticity of the turbulence and 
$\tau \sim (k_{\rm f} u_{\rm rms})^{-1}$ is the eddy correlation time scale. 
The prefactor $\epsilon_{\rm f}$, usually of order 0.1 or less is used 
since $({\vec{u}.\vec{\omega}})_{\rm rms} < u_{\rm rms}\omega_{\rm rms}$.
The case $\epsilon_{\rm f}=1$ 
means the flow is maximally helical. These approximations give us an 
estimate of $\alpha_0$ in terms of eddy diffusivity $\eta_{\rm t}$ and 
forcing scale $k_{\rm f}$ as,  
\begin{displaymath}
\alpha_0 = \epsilon_{\rm f}\frac{\tau k_{\rm f} u_{\rm rms}^2}{3} = \epsilon_{\rm f}\eta_{\rm t} k_{\rm f}.
\end{displaymath}
We would consider $k_{\rm f}$ rather than $\alpha_0$ as a free parameter in the 
model apart from $\eta_{\rm t}$. Assuming equipartition between magnetic energy 
and the turbulent energy, we also calculate an equipartition magnetic 
field $B_{\rm eq}$ as,
\begin{displaymath}
\beq = (4\pi\rho)^{1/2}u_{\rm rms}= (4\pi\rho)^{1/2} 3\eta_{\rm t} k_{\rm f}.
\end{displaymath}
For algebraic quenching we consider the following form for kinematic 
$\alpha_{\rm K}$ given by,
\begin{equation}
\label{eq:alphak}
\alpha_{\rm K}(r)=\frac{0.5\epsilon_{\rm f}\eta_{\rm t} k_{\rm f}
\Theta^+(r,r_a,d_a)
\cos\theta\sin^q\theta}{1+g_{\alpha}\ob{B}^2/\beq^2},
\end{equation}
where $g_{\alpha}$ is a non-dimensional coefficient equal to 1 or $R_{\rm m}$ 
depending on the assumed form of algebraic quenching in the models and
$q=0$ unless  
given. Even though the helical turbulence pervades almost the entire 
convection zone, we take $r_a = 0.77\Rs$ and $d_a=0.015\Rs$ so that we can 
have a large separation between the shear and turbulent layer. 
Consequently we consider a differential rotation profile like that 
in the high latitude tachocline of the Sun given by,
\begin{equation}
\label{eq:omega}
\Omega(r)=-\Omega_0\Theta^+(r,r_w,d_w),
\end{equation} 
where $\Omega_0=14$nHz, $r_w=0.68\Rs$ and $d_w=0.015\Rs$. The radial 
profiles of $\eta_{\rm t}$, $\alpha$ and $\pa \Omega/\pa r$ are plotted as a 
function of fractional radius $r/\Rs$ in Fig.~\ref{fig:profiles}. 
The region of strong radial shear is separated from the region of 
helical turbulence and the diffusivity has a strong gradient at a radius 
lying between the layers of finite strong shear and turbulent $\alpha$. 
The reason of the same is to decrease the time period $T_{\rm cyl}$ of the 
oscillatory dynamos to a reasonably small fraction of the diffusion time $t_{\rm diff}$. Our aim is to solve the induction equations coupled with yet 
another equation for the evolution of $\alpha$-effect, the formulation of 
which is described in \S2.1.

   \begin{figure}
   \label{fig:profiles}
   \centering
   \includegraphics[width=0.5\textwidth]{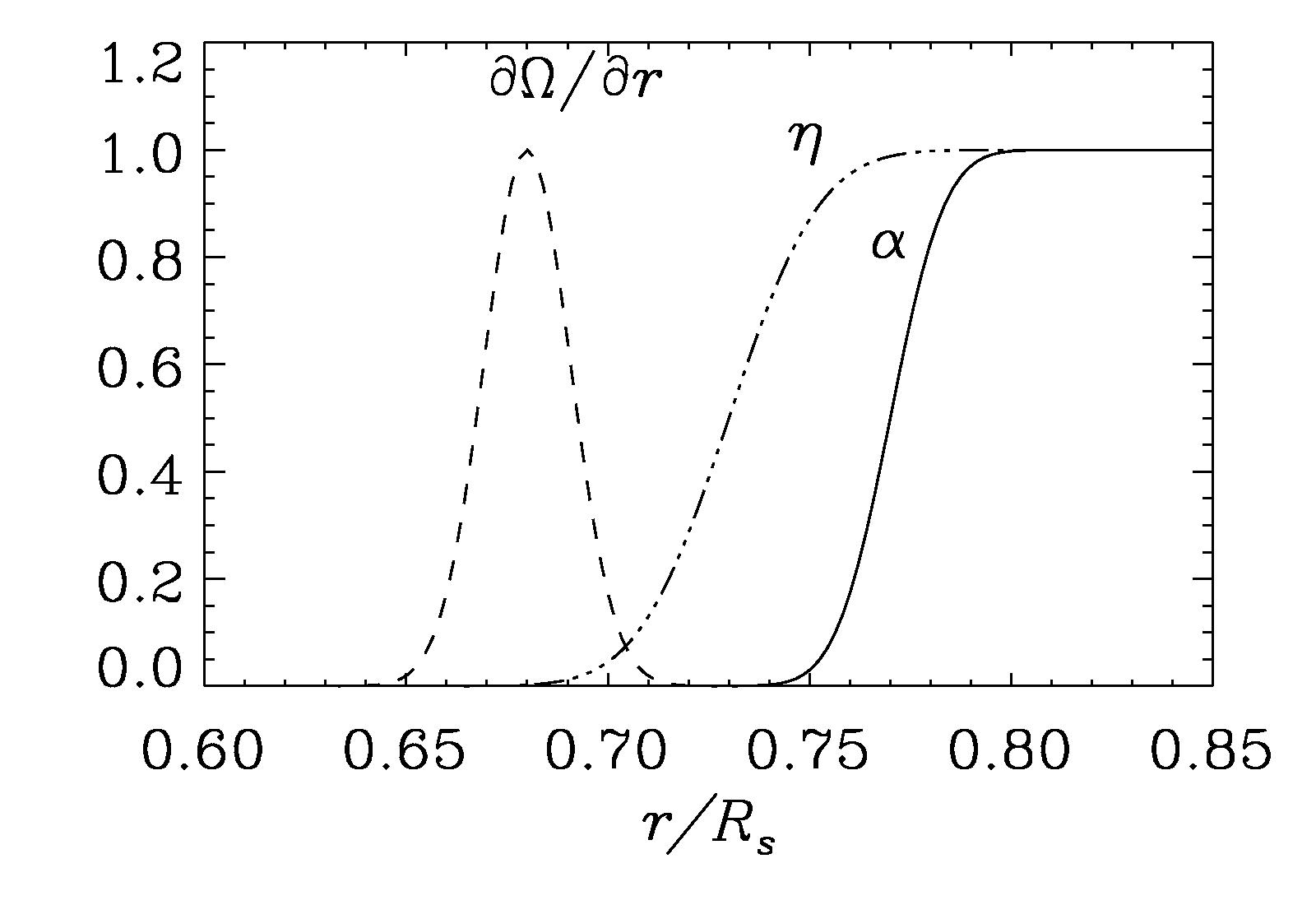}
   \caption{\label{fig:profiles} Profiles of radial shear $\pa \Omega/ \pa r$ (nHz cm$^{-1}$), $\alpha$ (cm s$^{-1}$) and 
           $\eta$ ($10^{12}$ cm$^2$ s$^{-1}$) as a function of fractional solar radius.}%
    \end{figure}
\subsection{Dynamical $\alpha$ quenching}
It was first shown by Pouquet et al.\ (1976) that the turbulent $\alpha$
effect is modified due to the generation of small-scale helicity in the
way given by Eq.~(\ref{eq:alpha}) below.
The second term is sometimes referred to as the magnetic $\alpha$-effect.
\begin{equation}
\label{eq:alpha}
\alpha=\alpha_{\rm K}+\alpha_{\rm M}=-\frac{\tau}{3}\left(\overline{\vec\omega\cdot\vec{u}} - \rho^{-1}\overline{\vec{j\cdot b}}\right),
\end{equation}
where $\vec\omega$, $\vec{u}$, $\vec{j}$, $\vec{b}$ denote the
fluctuating component of the  
vorticity, velocity, current and magnetic field in the plasma. It is 
possible to write an equation for the evolution of the magnetic part of 
$\alpha$ or $\alpha_{\rm M}$ from the equation for evolution of the
small-scale magnetic helicity density $h_{\rm f}=\overline{\vec{a\cdot
    b}}$ using the relation, 
\begin{equation}
\label{eq:rel}
\alpha_{\rm M} = \frac{\eta_{\rm t} k_{\rm f}^2}{B_{\rm eq}^2}h_{\rm f}.
\end{equation}
However the equation for $\overline{\vec{a\cdot b}}$ will be
gauge-dependent and  
it makes sense only to write an equation for the volume averaged quantity 
in order to avoid dependence on specific gauge (Blackman \& Brandenburg 2002).
Our dynamo equations are independent of any gauge since we solve 
for the magnetic potential component $A_{\phi}$ with an axisymmetric
constraint. It is important for us that the equation for $\alpha_{\rm M}$ is also
gauge independent. Subramanian \& Brandenburg (2006) used the Gauss linking 
formula for the expression for $h_{\rm f}$ and wrote an equation independent of 
the gauge for the magnetic helicity density under the assumption that 
the correlation length for all the fluctuating variables remain small 
compared to the system size at all times. Using Eq.~(\ref{eq:rel}) we write 
the same equation in terms of $\alpha_{\rm M}$,
\begin{equation}
\label{eq:alphaeq}
\frac{\pa \alpha_{\rm M}}{\pa t} = -2\eta_{\rm t} k_{\rm f}^2
\left(\frac{\overline{\vec{\mathcal{E}}}\cdot\overline{\vec{B}}}{B_{\rm eq}^2}
+\frac{\alpha_{\rm M}}{R_{\rm m}}\right)-\vec\nabla\cdot \overline{\vec{F}}_{\alpha},
\end{equation}
where $\vec{\mathcal{\overline{E}}}$ and $\overline{\vec{B}}$ are the
mean field EMF and the mean magnetic field.
The flux $\overline{\vec{F}}_{\alpha}$ consists of individual components, e.g.,
advection due to the mean flow, Vishniac--Cho fluxes
(Vishniac \& Cho 2001), effects of mean shear, diffusive fluxes, etc.
In this paper we have put $\overline{\vec{F}}_{\alpha}=0$ unless mentioned otherwise.

The decay time in Eq.~(\ref{eq:alphaeq}) is $t_{\alpha}=R_{\rm m}/\eta_{\rm t} k_{\rm f}^2=
4.55\times10^{-3} R_{\rm m} t_{\rm diff}$.
It should be noted that we use $g_{\alpha}=0$ in Eq.~(\ref{eq:alphak})
whenever we employ the dynamical quenching equation,
because dynamical quenching is usually more important.

\subsection{Flux transport Babcock-Leighton dynamo}
Axisymmetric mean field solar dynamo models including meridional circulation 
and Babcock-Leighton $\alpha$ effect have been studied extensively by 
several authors (Dikpati \& Charbonneau 1999; Chatterjee et al.\ 2004;
Guerrero \& Dal Pino 2008, and references therein). 
These models have now 
reached a stage where they are able to reproduce the butterfly 
diagram and the correct phase between the polar fields and the 
toroidal fields. In this section we will use a Babcock-Leighton (BL) $\alpha$ 
along with an analytical  meridional circulation (MC) which is poleward at the 
surface with a maximum amplitude of $u_0=20$ m s$^{-1}$ and the expression for which is
given by van Ballegooijen \& Choudhuri (1988). For completeness we provide the expressions for the radial and the latitudinal components of the meridional flow, $\vec{u}_p$ here.
\begin{equation}
u_r=u_0\left(\frac{\Rs}{r}\right)^2 \zeta \left(-\frac{2}{3}+
\frac{c_{s1}}{2}{\zeta}^{1/2}
-\frac{4c_{s2}}{9}{\zeta}^{3/4}\right)
     (2 \cos^2\!\theta-\sin^2\!\theta),
\end{equation}
\begin{equation}
u_{\theta}=u_0\left(\frac{\Rs}{r}\right)^3 (-1+c_{s1}\zeta^{1/2}-
c_{s2}\zeta^{3/4})\sin\theta \cos\theta,
\end{equation}
where $\zeta=\Rs/r-1$, $r_b=0.71\Rs$, $\zeta_b=\Rs/r_b-1$, $c_{s1}=4\zeta_b^{-1/2}$ and $c_{s2}=3\zeta_b^{-3/4}$.
It should be mentioned that, unlike in flux transport dynamo models, the
meridional circulation does not reverse the direction of propagation of the 
dynamo wave in interface dynamo models as long as the meridional circulation
is confined within the convection zone (Petrovay \& Kerekes 2004).
We solve this model along with the equation for dynamical $\alpha$ 
quenching described in Sect.~2.1. The fluxes in Eq.~(\ref{eq:alphaeq}) are 
now given by, 
\begin{equation}
\overline{\vec{F}}_{\alpha} = \alpha_{\rm M}{\vec{u}}_{\rm p}-\vec\nabla\cdot (\kappa\vec\nabla\alpha_{\rm M}),
\end{equation}
where $\kappa$ is the diffusion 
coefficient for $\alpha_{\rm M}$ taken to be $\kappa_0\eta(r)$. 
It may be remembered that the $\alpha_{\rm K}$ is now not due 
to the helical turbulence in the bulk of the convection zone, but due
to a phenomenological BL $\alpha$ where the poloidal field is produced from
the toroidal field due to decay of tilted bipolar active regions. The analytical
expression for $\alpha_{\rm K}$ is given by
\begin{eqnarray}
\label{eq:blalpha}
\alpha_{\rm K}=\frac{1}{4}\alpha_{\rm BL}
\Theta^{+}(r, 0.95\Rs, d)\;
\Theta^{-}(r, \Rs, d)\cos\theta\sin^2\theta
\end{eqnarray}
with $d=0.015\Rs$.
The BL $\alpha$ is assumed to be concentrated only in the upper 0.05\% of the
convection zone.
The turbulent diffusivity has the same profile as in Eq.~(\ref{eq:eta}) but with 
$\eta_{\rm t}=2\times10^{11}$ cm s$^{-1}$ and $r_e=0.7\Rs$. The shear is still radial and given by 
Eq.~(\ref{eq:omega}) with $r_w=0.7\Rs$.

Our computational domain is defined to be the region confined by $0 \le \theta \le \pi$ and
$0.55\Rs\le r \le \Rs$. Unless otherwise stated, the boundary conditions for $A_{\phi}$ are given by a potential field condition at the surface 
(Dikpati \& Choudhuri 1994) and $A_{\phi}=0$ at the poles. We have also 
performed some calculations with the vertical field condition 
at the top boundary, which means that $B_{\theta}=B_{\phi}=0$.
At the bottom we use the perfect conductor boundary condition of 
Jouve et al.\ (2008) with $A_{\phi} = \pa(rB_{\phi})/\pa r=0$. However
a more realistic perfect conductor boundary condition in our
opinion would be $\pa (r B_{\theta})/\pa r = \pa(rB_{\phi})/\pa r=0$. 
Also $B_{\phi}=0$ on all other boundaries.  The equation for 
$\alpha_{\rm M}$ is an initial value problem for $F_{\alpha}=0$. For finite 
fluxes we have also set $\alpha_{\rm M}=0$ at all boundaries. We have checked 
that the results are not very sensitive to the different boundary conditions
given above mainly because the boundaries are far removed from the dynamo 
region.
%
\section{Results}
\subsection{Magnetic field properties without helicity fluxes}
\begin{figure}
   \includegraphics[width=0.5\textwidth]{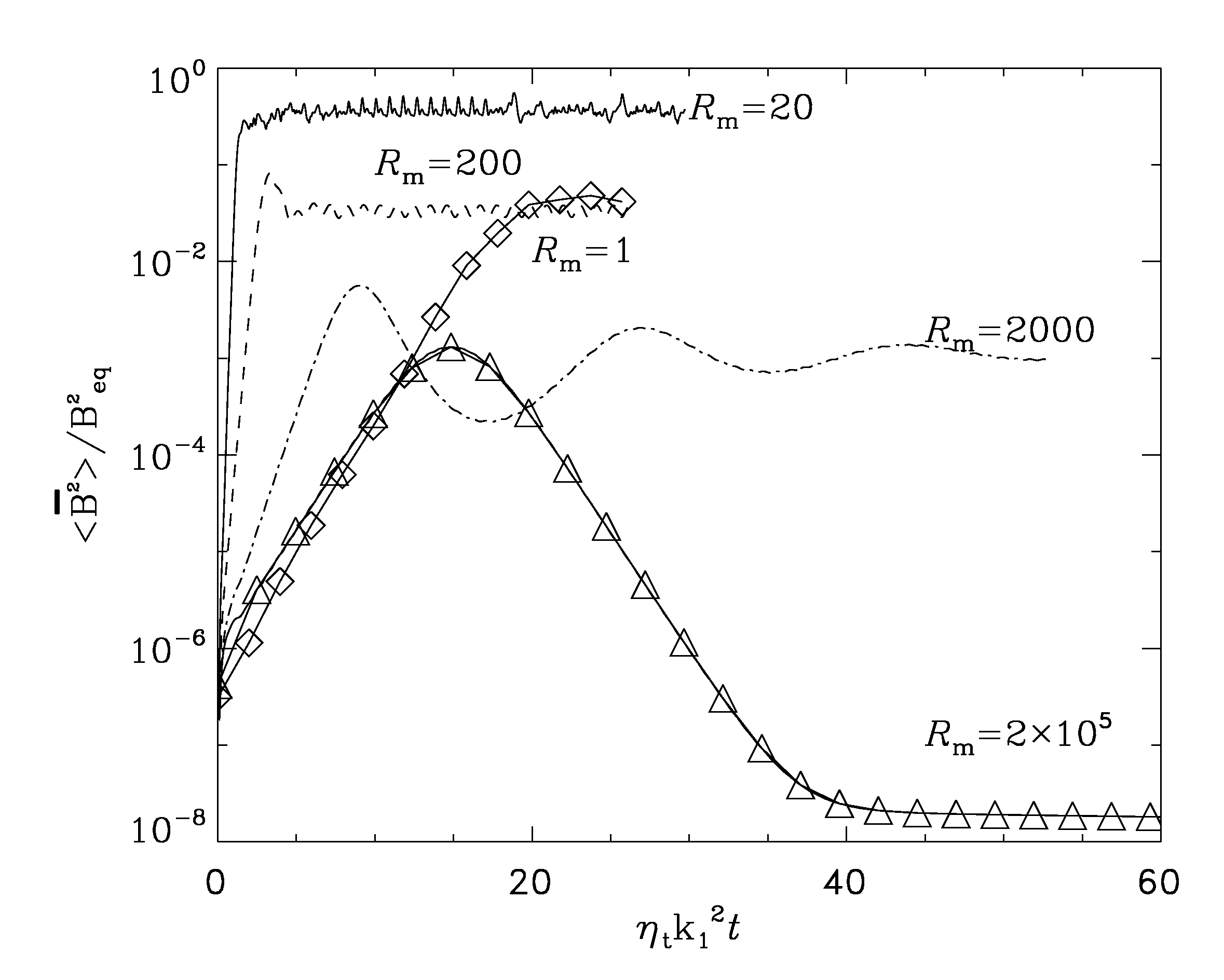}
      \caption{Magnetic energy in the domain scaled with the equipartition energy for $R_{\rm m}=1$ (diamond+line), $R_{\rm m}=20$ (solid), $R_{\rm m}=200$ (dashed), $R_{\rm m}=2\times10^3$(dashed-dotted), $R_{\rm m}=2\times10^5$ (triangles+line) for $\alpha_0=0.08\eta_{\rm t} k_{\rm f}$ for the $R_{\rm m}$ indicated in the figure.
              }
         \label{fig:energy1}
   \end{figure}

In order to study the $R_{\rm m}$ dependence of the saturation magnetic
field in the two layered dynamo with diffusive coupling we keep all
the dynamo parameters the same for all the runs and change $\eta_r$
from $2\times10^5$ cm$^2$ s$^{-1}$ to $2\times10^{10}$ cm$^2$ s$^{-1}$
while keeping $\eta_{\rm t}$
fixed at $4\times10^{10}$ cm$^2$ s$^{-1}$. It may also be noted that 
the time period of the dynamo models ($T_{\rm cyl}$) is fairly 
independent of the magnetic Reynolds number. We show the magnetic 
energies as a function of time for the nonlinear system with $\alpha_0=0.08\eta_{\rm t} k_{\rm f}$ for a range of magnetic Reynolds 
numbers in Fig.~\ref{fig:energy1}. The strong $R_{\rm m}$ dependence 
which is reminiscent of catastrophic quenching in all astrophysical 
dynamos can be easily discerned from 
Fig.~\ref{fig:energy1}. It is interesting that the saturation energy 
of the $R_{\rm m}=1$ model is lower than that of the $R_{\rm m}=20$ case. The dynamo model 
may be highly dissipative at very low magnetic Reynolds numbers.  

   \begin{figure}
   \includegraphics[width=0.5\textwidth]{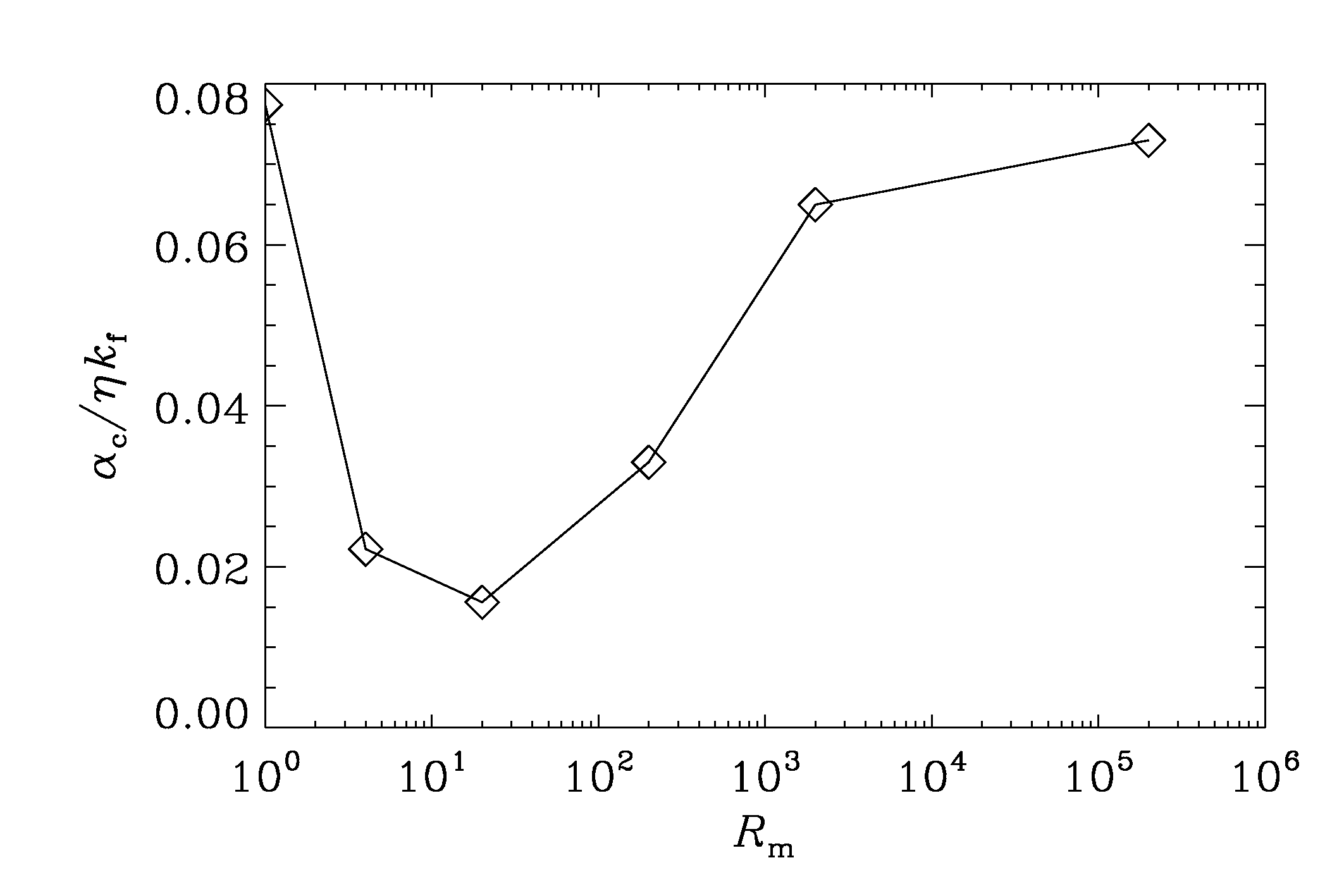}
      \caption{Critical $\alpha$ in terms of a fraction of $\eta_{\rm t} k_{\rm f}$ as a function of magnetic Reynolds number $R_{\rm m}$ for the interface dynamo model of Fig.~\ref{fig:profiles}. 
              }
         \label{fig:crit}
   \end{figure}

The slopes of the volume averaged energy are also very different in the 
kinematic phase, which means that the critical dynamo numbers also depend 
on $R_{\rm m}$. To be able to correctly compare the dynamo models for different $R_{\rm m}$, 
it is first important to calculate the critical value of $\alpha_0$, denoted by $\alpha_c$ for each model. Such a plot is shown in Fig.~\ref{fig:crit}. 
From this figure we can conclude that this dynamo model is most efficient 
near $R_{\rm m}=20$. A similar variation of $\alpha_c$ with the ratio $\eta_{\rm t}/\eta_r$ 
was obtained analytically for interface dynamos by
MacGregor \& Charbonneau (1997; see their Fig.~5A).
We now set $\alpha_0=2\alpha_c$, corresponding to the 
$R_{\rm m}$ of each model, and repeat our calculations. We shall now use this value
of $\alpha$ for the rest of the paper. 
The saturation energy decreases monotonically as a function of magnetic 
Reynolds number as shown in Fig.~\ref{fig:satrm}.
For $R_{\rm m} =2\times10^5$, 
the code has to be run for 500 $t_{\rm diff}$ before the dynamo field starts 
becoming 'strong' again for the case with $\alpha_0=2\alpha_c$. 
Due to long computational times involved in this exercise we have not 
continued the calculation beyond 60 $t_{\rm diff}$.
Hence, the determination of saturation magnetic energy may be inaccurate
for $R_{\rm m}=2\times10^5$. 
Compare this with the case of a simple algebraic quenching of the 
form given in Eq.~(\ref{eq:alphak}) with $g_{\alpha}=1$. The slopes in the kinematic phase are now almost similar for all $R_{\rm m}$ within the error 
in the numerical determination of the critical $\alpha_c$. 
For $g_{\alpha}=R_{\rm m}$, the algebraically and dynamically quenched
$\alpha$ effects seem to give similar dependences on $R_{\rm m}$.
It may occur that the two source regions may not 
be spatially separated, so we repeat our calculations with the $\alpha$ 
region at $r_a=0.87\Rs$ instead of $0.77\Rs$ and obtain the same slope in 
the relation of the volume averaged magnetic energy on $R_{\rm m}$ as in Fig.~\ref{fig:satrm}. We also verify from the profiles of field components 
at two different latitudes, as shown in Fig.~\ref{fig:bphi}, that the region
of strong toroidal field $B_{\phi}$ is different from the layer where poloidal
fields are produced by the $\alpha$ effect.

       \begin{figure}
      \includegraphics[width=0.5\textwidth]{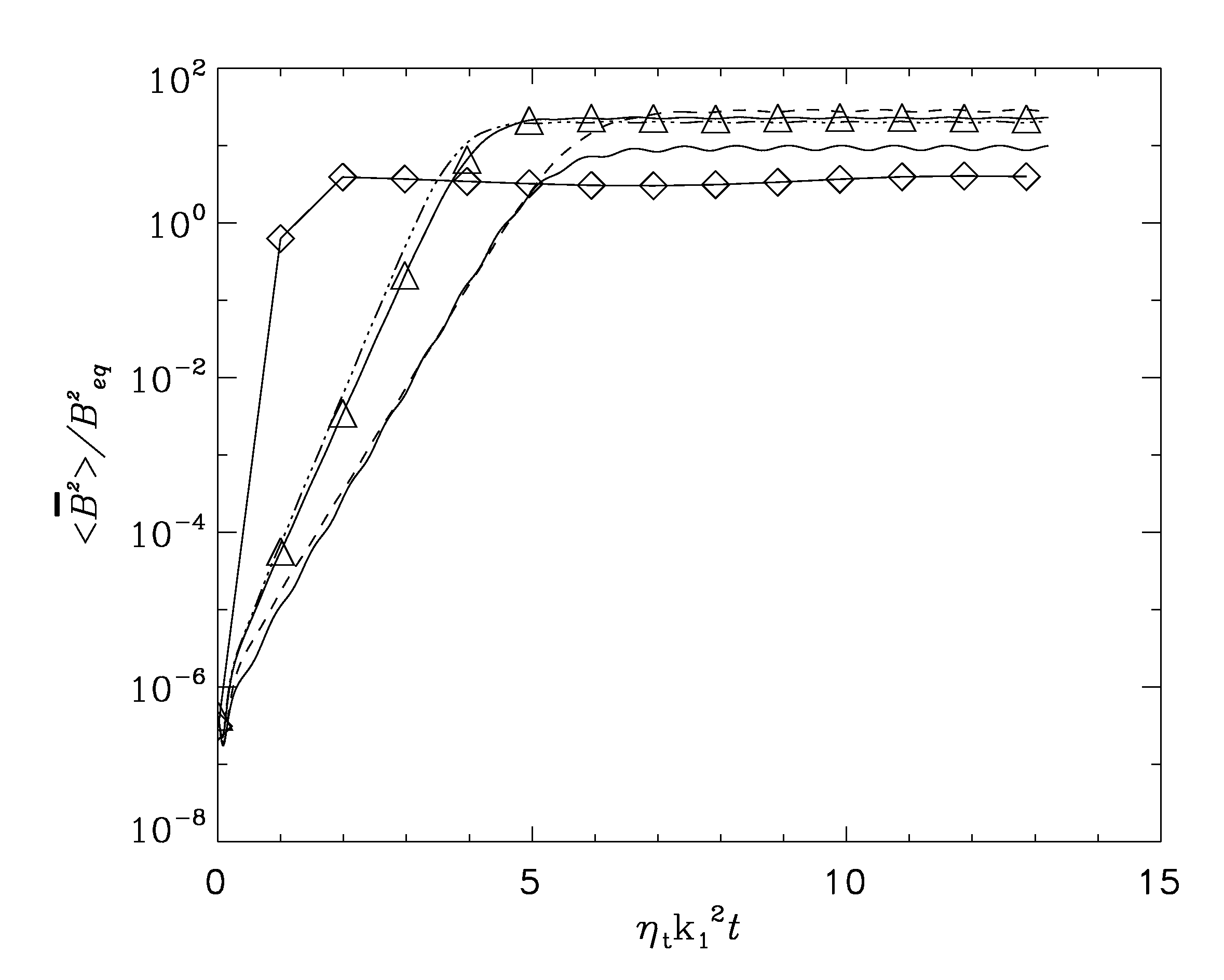}
      \caption{Magnetic energy in the domain scaled with the equipartition energy for $R_{\rm m}=1$ (diamond+line), $R_{\rm m}=20$ (solid), $R_{\rm m}=200$ (dashed), $R_{\rm m}=2\times10^3$ and $R_{\rm m}=2\times10^5$ 
(triangles) with algebraic quenching for $g_{\alpha}=1$
              }
         \label{fig:energy3}
   \end{figure}
   
       \begin{figure}
      \includegraphics[width=0.5\textwidth]{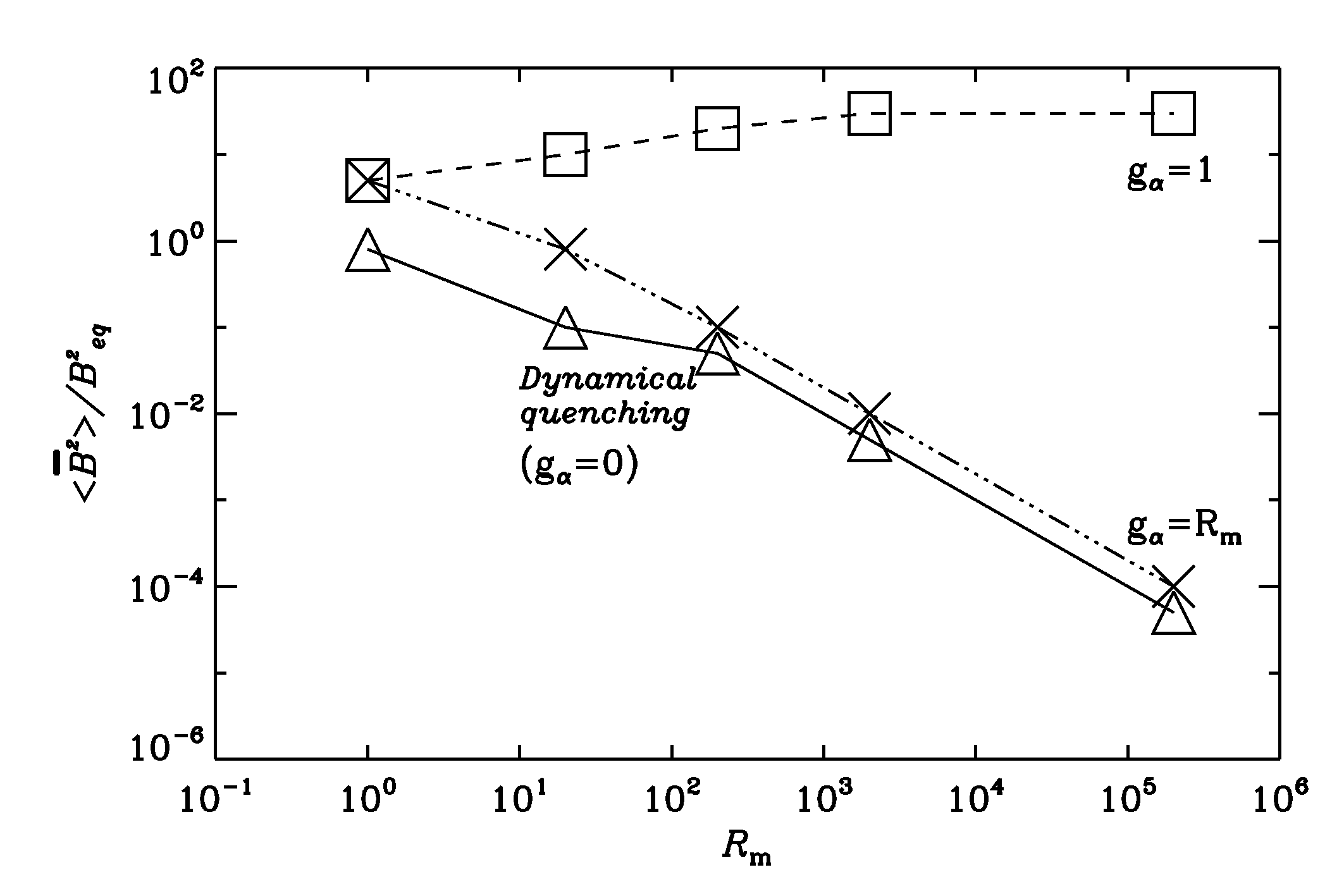}
      \caption{Volume averaged magnetic energy scaled with the equipartition energy  in the saturation phase as a function of $R_{\rm m}$ for dynamical $\alpha$ quenching (triangles +solid) and algebraic quenching with $g_{\alpha}=1$ (squares + dashed) and with $g_{\alpha}=R_{\rm m}$ (cross + dashed-dotted).
              }
         \label{fig:satrm}
   \end{figure}
   
For the solutions with dynamical $\alpha$ effect, it may be concluded from
the butterfly diagrams of Fig.~\ref{fig:alp_bfly}
that the small-scale current helicity $\alpha_{\rm M}$ is predominantly negative
(positive) in the Northern (Southern) hemisphere.
The nature of the saturation curves of the magnetic energy is strongly
governed by the ratio of $t_{\alpha}$ and $T_{\rm cyl}$.
For $R_{\rm m}=20$, $t_{\alpha} \ll T_{\rm cyl}=0.85t_{\rm diff}$ and so there are 
strong oscillations in the butterfly diagram for $\alpha_{\rm M}$,
as shown in Fig.~\ref{fig:alp_bfly}a, whereas for $R_{\rm m}=200$,
$t_{\alpha} \sim T_{\rm cyl}$
the amplitude of oscillations is weak because the $\alpha_{\rm M}$ decays at 
the same rate at which it is produced due to the effect of the oscillatory 
source term $\vec{\mathcal{E}}\cdot\overline{\vec{B}}$; see Fig.~\ref{fig:alp_bfly}b. 
Similarly for $R_{\rm m}=2\times10^3$, the decay time $t_{\alpha} \gg T_{\rm cyl}$ 
and so the system of equations is overdamped as can be seen from the 
saturation curve (dashed dotted line) in Fig.~\ref{fig:energy1}b where 
there are amplitude modulations of the magnetic field before it 
settles to a final saturation value. When the code is run longer, 
we start seeing changes in the parity after $t > 40 t_{\rm diff}$. 
However the magnetic energy and the dynamo period $T_{\rm cyl}$ remain
fairly constant even while the system fluctuates between symmetric 
and anti-symmetric parity at an irregular time interval 
(see Fig.~\ref{fig:parity}). This parity oscillation is absent in the 
corresponding models with algebraic quenching. 
      \begin{figure}
   \includegraphics[width=0.5\textwidth]{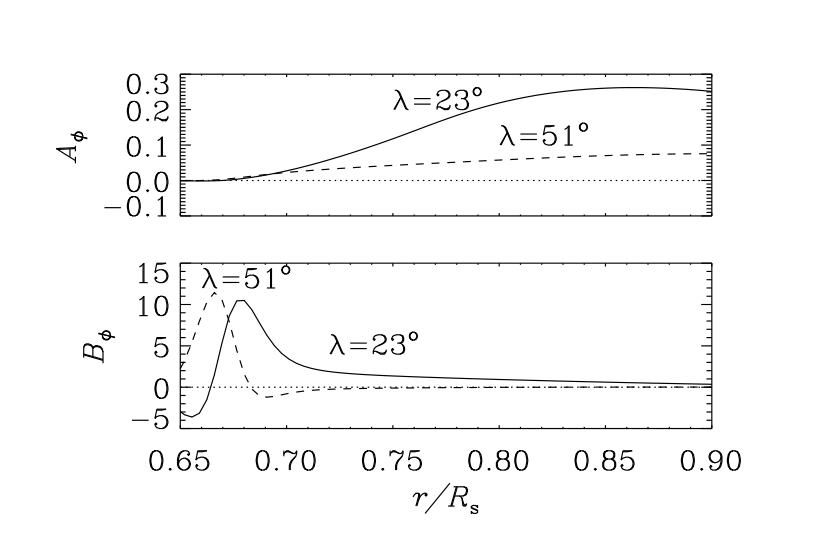}
      \caption{Radial profiles of $A_{\phi}$ and $B_{\phi}$ at two different latitudes ($\lambda$) in the saturated phase for $R_{\rm m} = 2\times 10^3$.
              }
         \label{fig:bphi}
   \end{figure}

\subsection{Secondary dynamo waves}

An interesting result emerges when we repeat our calculations with 
$\alpha_0=4\alpha_c$ instead of $2\alpha_c$ for $R_{\rm m}=20$.
The negative $\alpha_{\rm M}$ generated in the convectively unstable layer 
penetrates below 0.73$\Rs$ where $\alpha_{\rm K}=0$ and drives a secondary 
dynamo wave whose direction of propagation is poleward as compared to the 
primary dynamo wave propagating equatorward. This can be seen in the 
butterfly diagram of $B_{\phi}$ at 0.72$\Rs$ in Fig.~\ref{fig:supercrit}a.
Signature of the secondary dynamo can also be seen in the butterfly
diagram at 0.8$\Rs$.
Even though the secondary dynamo wave is energetically powered by the
kinematic part of the helical convection but the direction of propagation
is governed by the sign of $\alpha_{\rm K}+\alpha_{\rm M}$.
This may be compared with an $\alpha\Omega$ dynamo driven by a 
supercritical helicity flux (Vishniac \& Cho 2001).
This mechanism however requires finite initial magnetic field.
It may be recalled that we have done
calculations with an initial field $\sim 10^{-6} B_{\rm eq}$.
The difference compared to the case above is that the mean field dynamo
is not driven by supercritical Vishniac \& Cho fluxes, but it is
governed by a local generation of small-scale magnetic helicity.
We return to the issue of secondary dynamo waves driven by diffusive
magnetic helicity fluxes in Sect.~\ref{Diffusive}.

We also have not observed any 
evidence of chaotic behaviour in the range of magnetic Reynolds 
number $20 \le R_{\rm m} \le 2\times 10^5$ for supercritical 
$\alpha\leq 4\alpha_c$ in agreement with Covas et al.\ (1998).
However, if the $\alpha$ effect is highly supercritical, the dynamical
quenching formula for $\alpha_{\rm M}$ is insufficient for dynamo saturation,
and additional algebraic quenching terms must enter
(Kleeorin \& Rogachevskii 1999).

   \begin{figure}
   \includegraphics[width=0.5\textwidth]{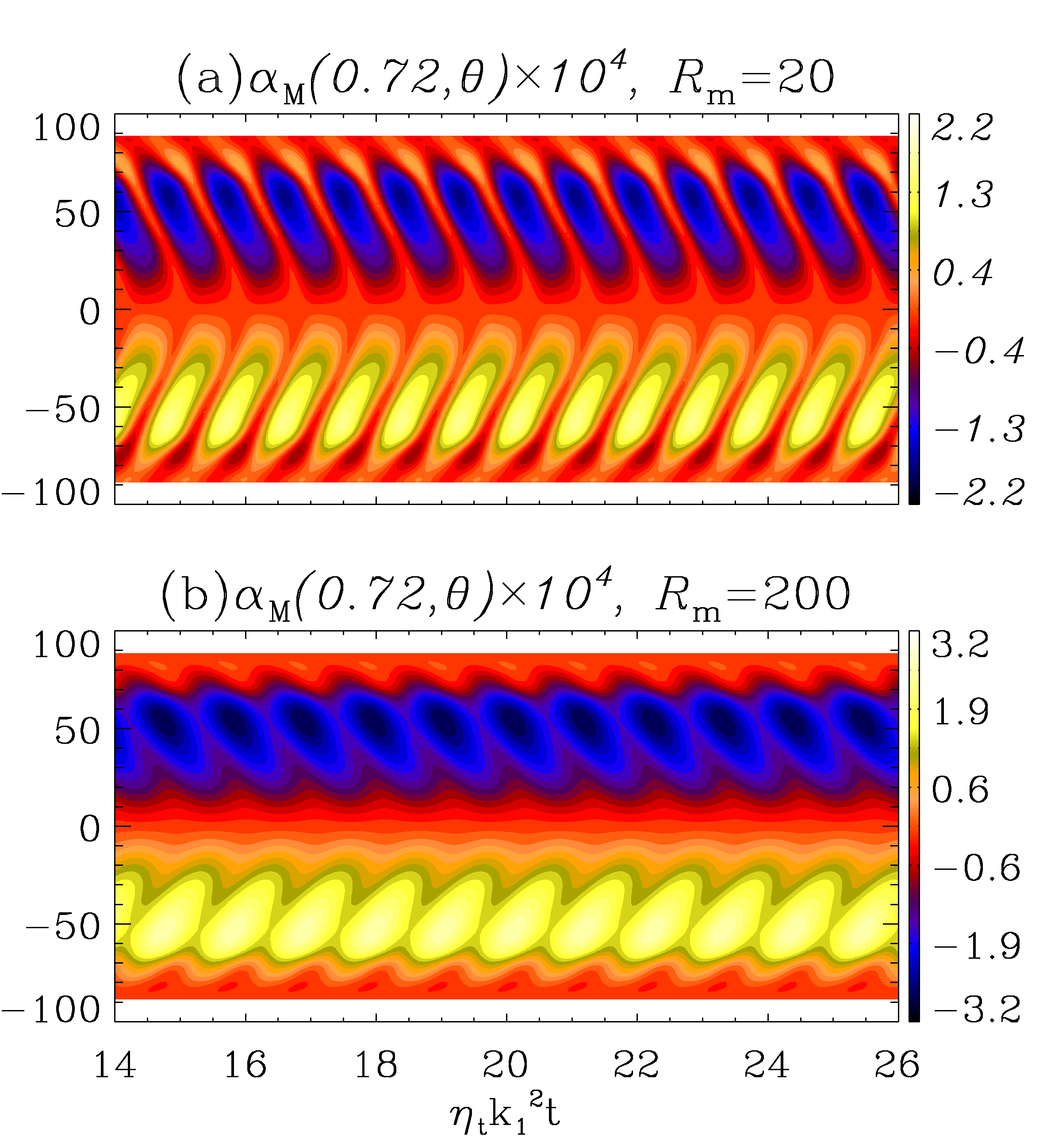}
      \caption{$\alpha_{\rm m}(0.72\Rs,\theta)$ as a function of diffusion time $\eta_{\rm t} k_1^2t$ for (a) $R_{\rm m}=20$ and (b) $R_{\rm m} = 200$.
              }
                 \label{fig:alp_bfly}
   \end{figure}
   
\subsection{Diffusive magnetic helicity fluxes}
\label{Diffusive}

Recently, Brandenburg et al.\ (2009) showed that catastrophic quenching
in one-dimensional $\alpha^2$ dynamos can be alleviated by introducing
a Fickian diffusive flux in Eq.~(\ref{eq:alphaeq}) given by
\begin{equation}
\overline{\vec{F}}_{\alpha} = -\kappa \vec\nabla\alpha_{\rm M}.
\end{equation}
There was an attempt to calculate the diffusion coefficient $\kappa$
from direct numerical simulations and it was found to be $\sim 0.3\eta_{\rm t}$
for $R_{\rm m} \sim 20$ (Mitra et al.\ 2010).
For $\kappa=0$, the saturation curves in Fig.~\ref{fig:energy1}b 
show that the $B_{\rm sat}$ goes through very low values for
$R_{\rm m} \sim 2\times 10^5$ and it takes very long to relax
to a steady amplitude. 
Next we introduce a diffusive 
flux with $\kappa(r) = \kappa_0\eta(r)$ in Fig.~\ref{fig:energy4} and obtain
$B_{\rm sat}\sim 0.1B_{\rm eq}$ and underdamped behaviour. However looking
carefully at the corresponding butterfly diagrams (Fig.~\ref{fig:diffflux}a,b,c,d) we find a poleward propagating mode due to radial diffusion of the $\alpha_{\rm M}$ 
into the stable layers which otherwise was not possible for a very high $\eta_{\rm t}/\eta_r$ ratio.
Figures~\ref{fig:diffflux}e,f show meridional snapshots of 
$\textrm{sign}(B_{\phi})(|B_{\phi}|/B_{\rm eq})^{1/2}$ and $\alpha_{\rm M}$ in order 
to get a clearer idea of the distribution of magnetic fields. 
The poleward propagating mode is now driven by supercritical diffusive helicity
fluxes, as opposed to supercritical Vishniac \& Cho fluxes
(see Brandenburg \& Subramanian 2005 for examples of such behaviour).
There exists a $\kappa_c \sim 10^{-5}$ for $\Rm=2\times 10^5$
such that the secondary dynamo fails to operate if $\kappa_0 < \kappa_c$ and the volume averaged magnetic energy decays eventually. It should be noted
that this threshold for $\kappa$ is highly dependent on $\Rm$. 
For instance $\Rm=2\times10^3$
and $\kappa_0=10^{-5}$ produces a dynamo with finite saturation magnetic field and dynamo wave propagation governed by $\alpha_M$ where as for 
$\kappa=10^{-4}\eta$, the dynamo shows a runaway growth. An interesting behaviour can be discerned from the butterfly diagram of the toroidal field  
for $\Rm=2\times 10^3$ and $\kappa_0=10^{-5}$ (Fig.\ref{fig:energy4}b,c). It appears that the behaviour of the dynamo is governed by competition between the poleward propagating mode and the equatorward propagating mode. The volume averaged energy (stars+line in Fig.\ref{fig:energy4}a) shows corresponding oscillations long after saturation at an period $\sim 5$ times the period of the equatorward propagating mode. 
It may be 
recalled that it is well established from direct numerical simulations of
$\alpha^2$ dynamos that a large-scale magnetic field is easily excited on
the scale of the system i.e., $k_1^{-1}$ for a large $k_{\rm f}/k_1$ ratio 
(Archontis, Dorch, Nordlund, 2003).
The length scale of the magnetic field in Figs.\ref{fig:energy4}b,c and Figs.~\ref{fig:diffflux}a,c,e
is comparable to $k_{\rm f}^{-1}$, which suggests that the degree
of scale-separation may have become insufficient to write the
electromotive force as a simple multiplication, as is done in the
expression $\overline{\vec{\mathcal{E}}} = \alpha \overline{\vec{B}} - \eta \overline{\vec{J}}$, and
that it may have become necessary to write it as a convolution, which
corresponds essentially to a low-pass filter
(see, e.g., Brandenburg et al.\ 2008).
However, we have not pursued this aspect any further.

      \begin{figure}
   \includegraphics[width=0.5\textwidth]{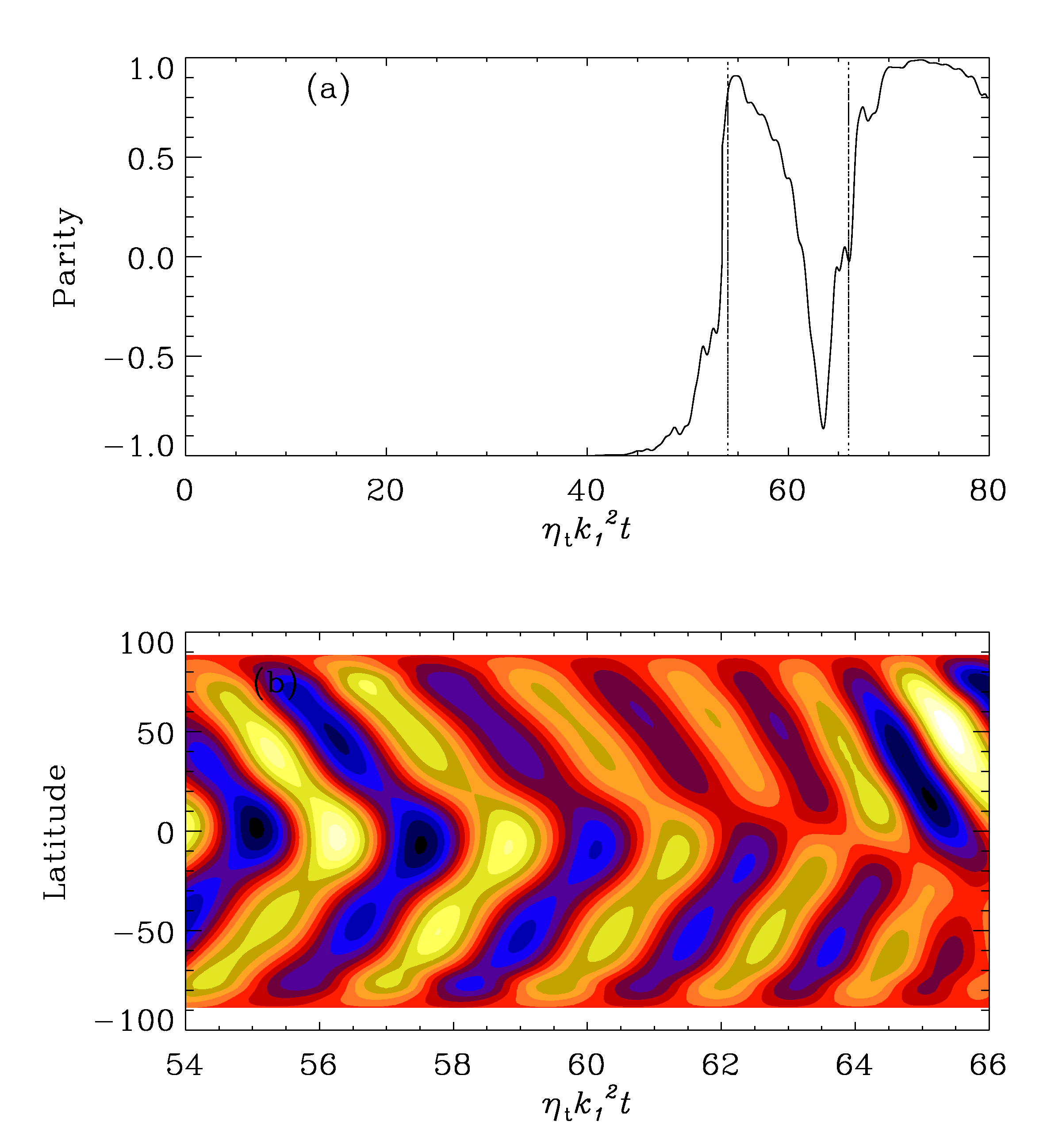}
      \caption{(a) Evolution of parity (purely dipolar $=-1$ and 
      purely quadrupolar $= +1$) for $\Rm=2\times10^3$. (b) A small part in the
      butterfly diagram indicated by dotted lines in (a) where parity is changing from quadrupolar to dipolar.
              }
         \label{fig:parity}
   \end{figure}
   
         \begin{figure}
   \includegraphics[width=0.5\textwidth]{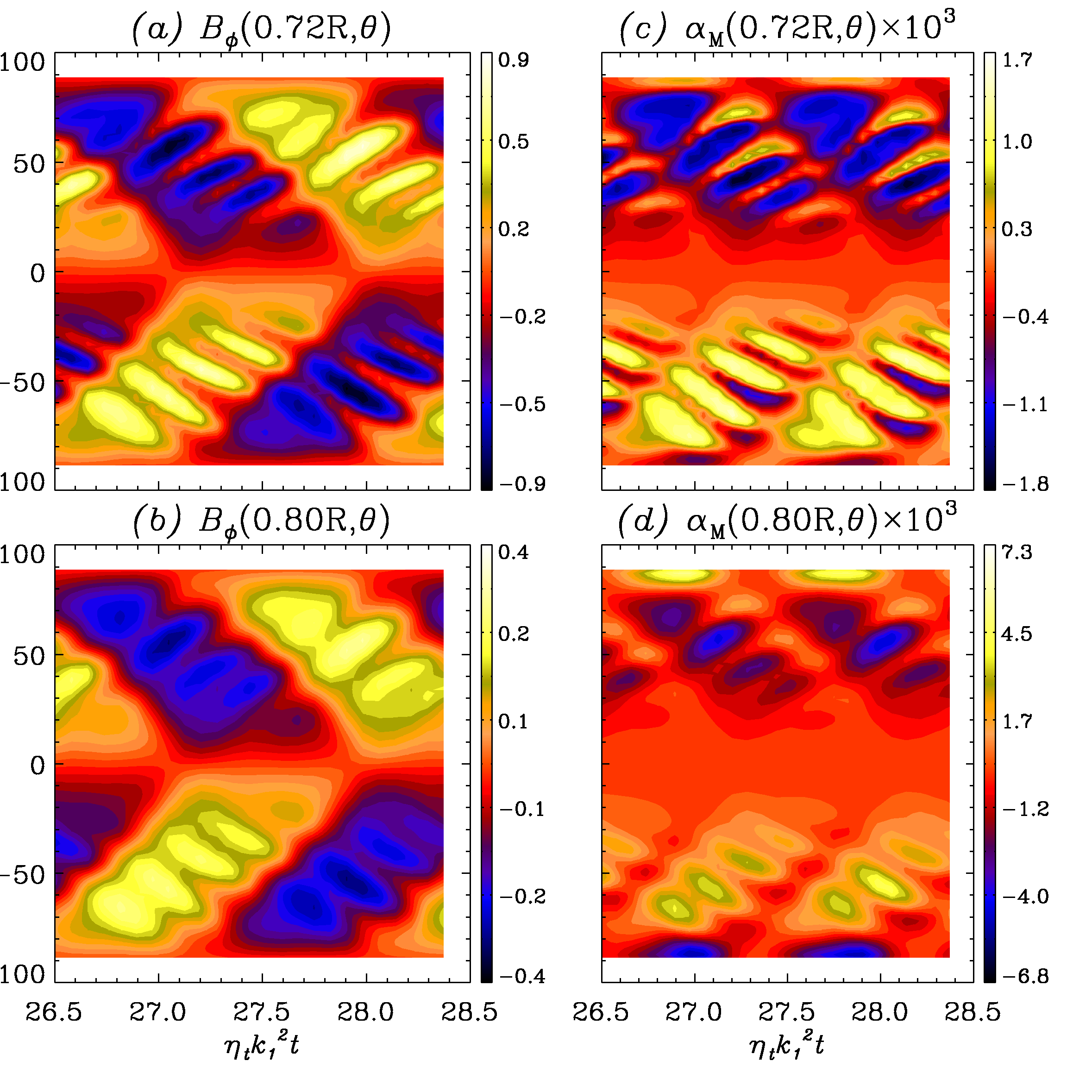}
      \caption{Butterfly diagrams of the toroidal field (a) and (c) and $\alpha_{\rm m}$ (b) and (d) with $\alpha=4\alpha_c$ for $R_{\rm m}=20$.
              }
         \label{fig:supercrit}
   \end{figure}
   
      \begin{figure}
            \label{fig:energy4}
      \includegraphics[width=0.5\textwidth]{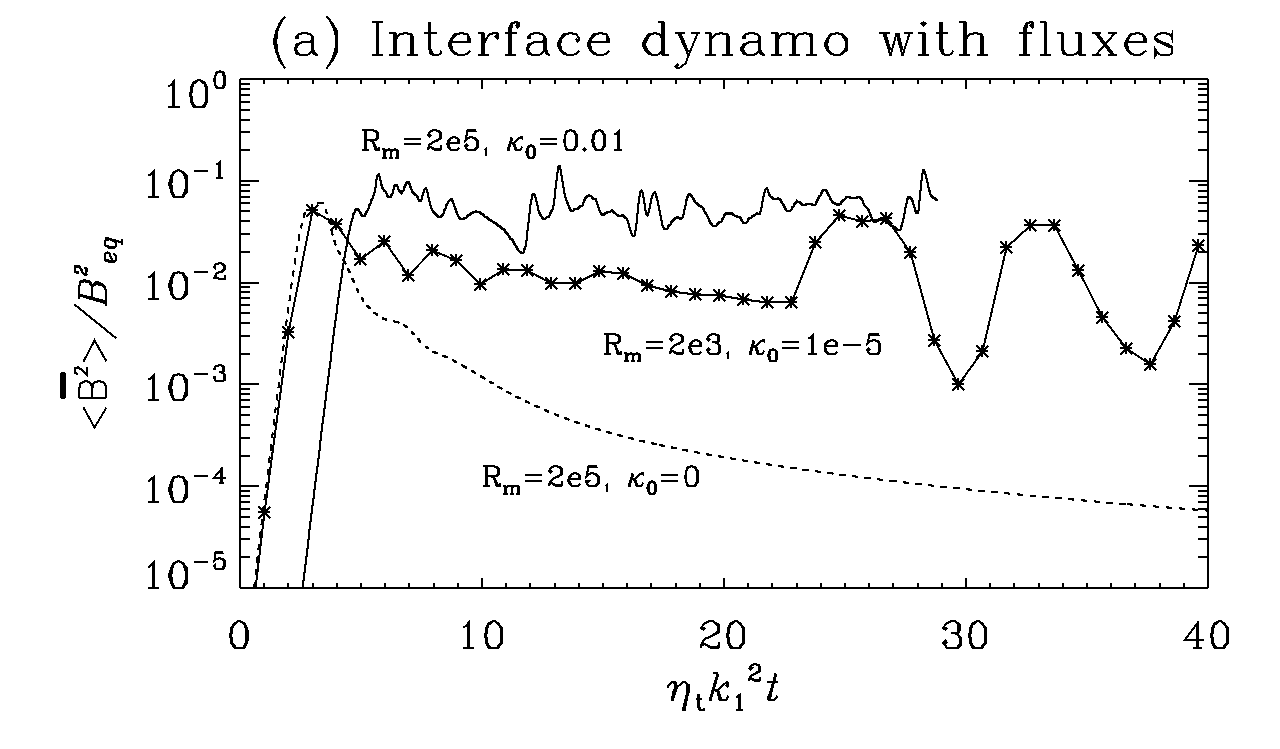}
      \includegraphics[width=0.5\textwidth]{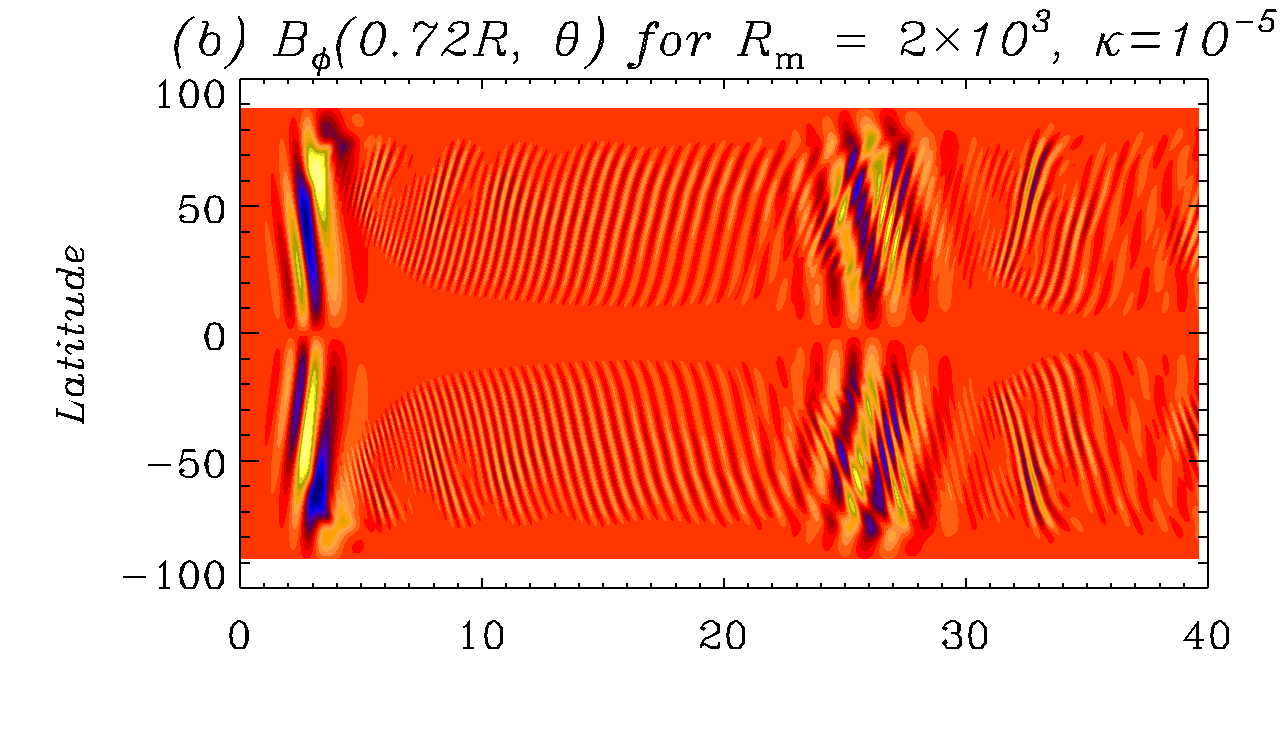}
          \includegraphics[width=0.5\textwidth]{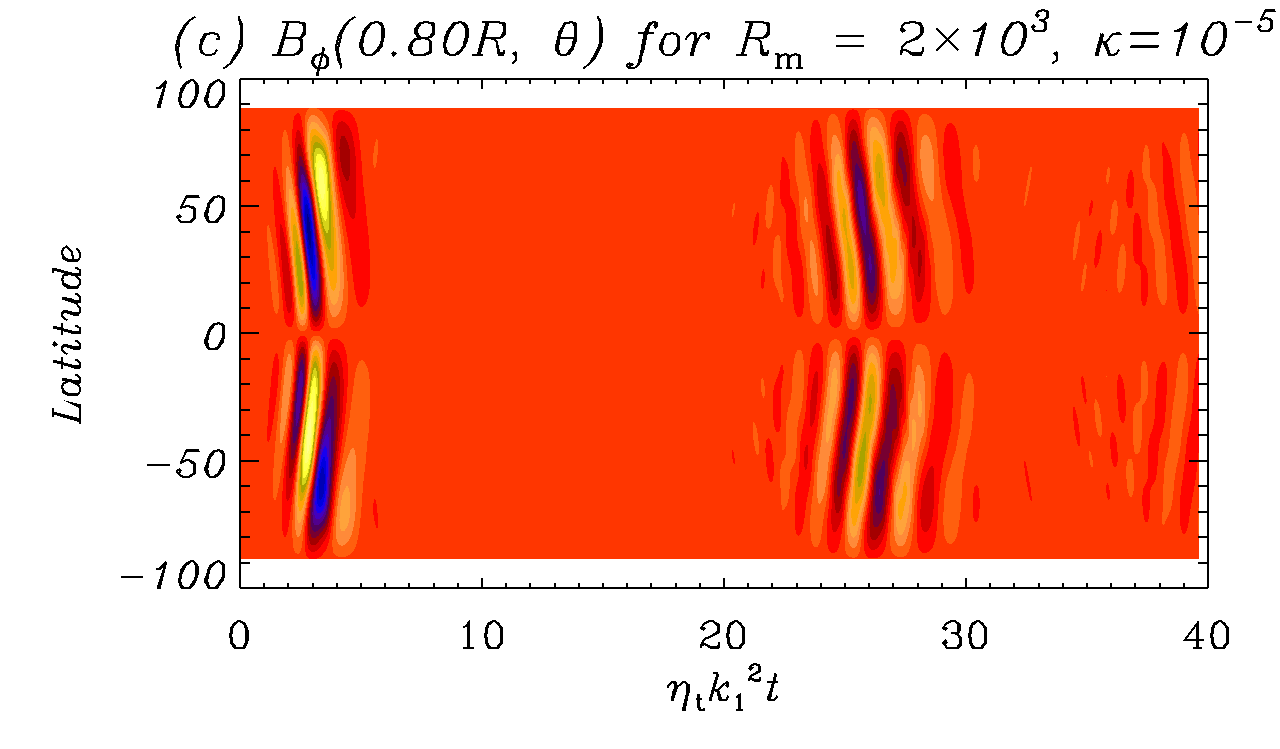}
      \caption{\label{fig:energy4} (a) Magnetic energy in the domain scaled with the equipartition energy for $R_{\rm m}=2\times10^3$  indicated in the figure 
      for the case of two layered dynamo of 
      Sect.~3.1 with dynamical $\alpha$ quenching with a diffusive flux 
      with $\kappa_0=10^{-5}$ (star+solid). The same for 
      $R_{\rm m}=2\times10^5$ and $\kappa_0=10^{-2}$ (solid).
       The saturation curve for 
      zero fluxes have been shown by the dashed line. (b) and (c) show butterfly 
      diagrams for the toroidal field at the depths indicated for
       $\Rm=2\times10^3$ and $\kappa_0=10^{-5}$.}
      \end{figure}

         \begin{figure}
      \includegraphics[width=0.5\textwidth]{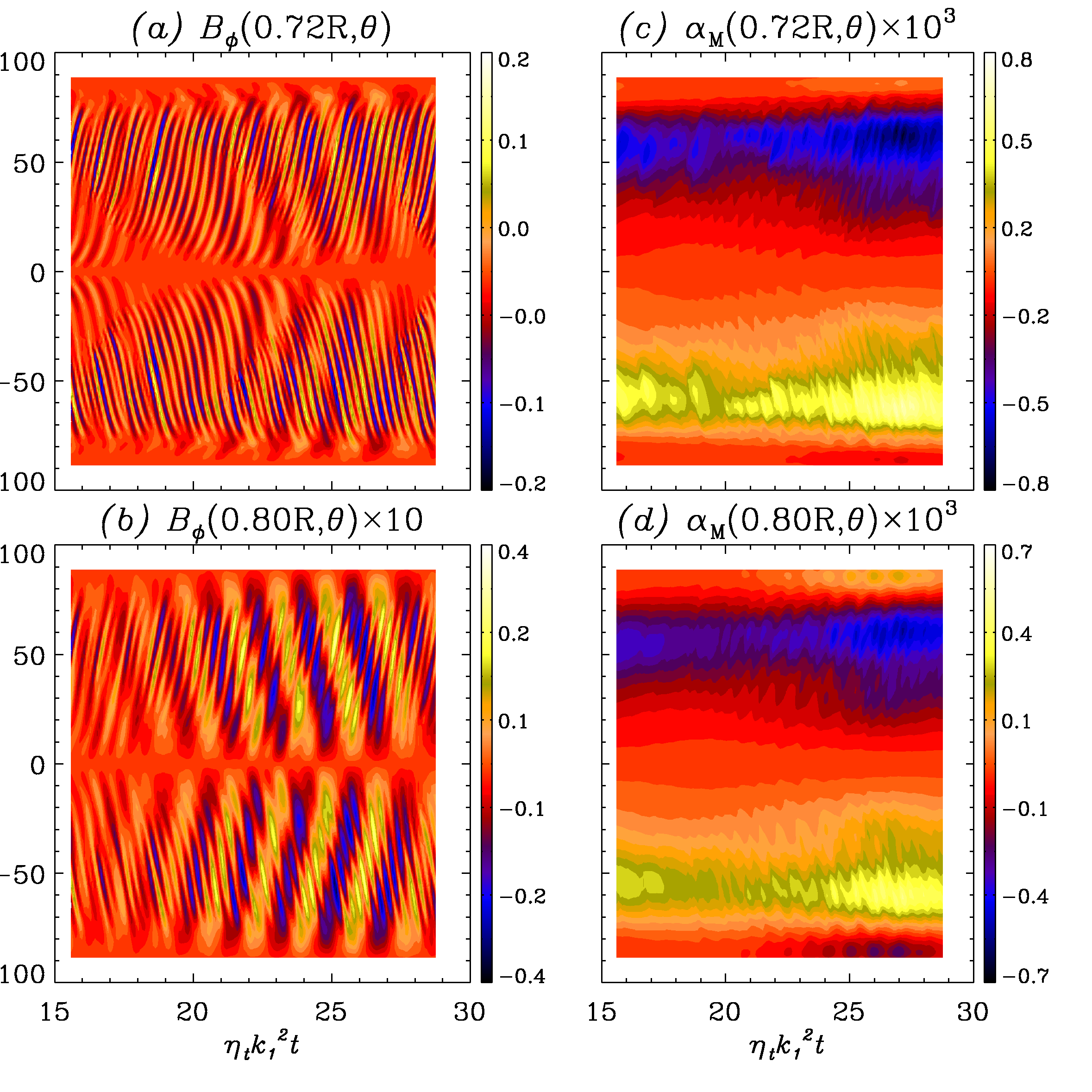}
      \includegraphics[width=0.5\textwidth]{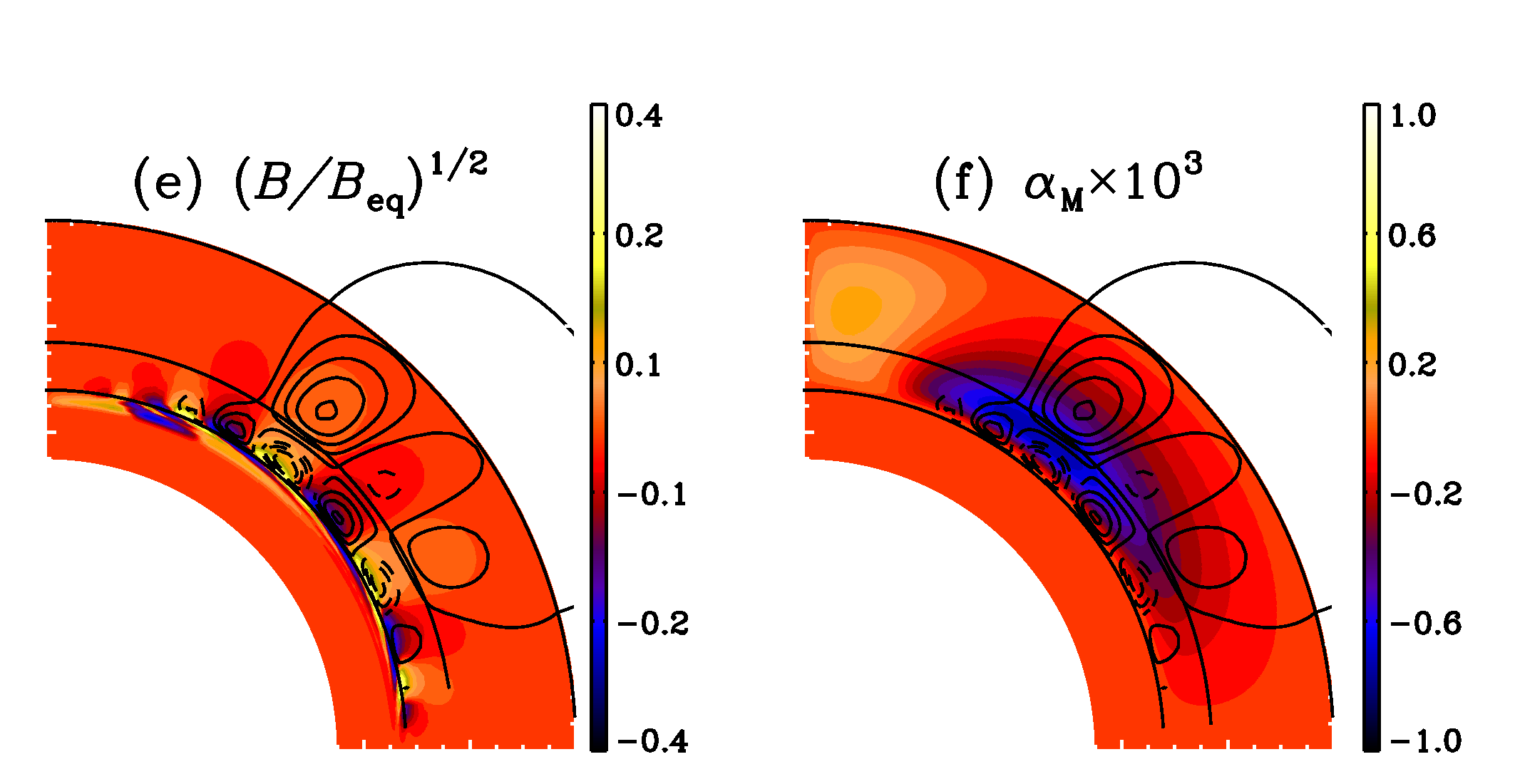}
      \caption{Butterfly diagrams of the toroidal field (a) and (c) and $\alpha_{\rm m}$ (b) and (d) with $\alpha=2\alpha_c$ for $R_{\rm m}=2\times10^5$ with a $\kappa_0=0.01\eta_{\rm t}$. Meridional snapshots of (e) $(B/B_{\rm eq})^{1/2}$ and (f) $\alpha_{\rm m}\times 10^3$ for the same case. 
              }
         \label{fig:diffflux}
   \end{figure}

\subsection{Flux transport Babcock-Leighton Dynamo}
Like in \S3.1 we find the critical $\alpha_{\rm BL}$ required to have a self 
excited dynamo. In this case $\alpha_c=5.1$ m s$^{-1}$ for $\Rm=2\times10^3$. 
We pursue the rest
of the calculations with $\alpha_{\rm BL}=6.0$ m s$^{-1}$ in order to avoid 
producing very large $\alpha_{\rm M}$ leading to secondary dynamos discussed in \S3.1.
We should emphasize that Eq.~(\ref{eq:alpha}) represents a first order 
correction to the $\alpha$ and should be treated with caution during its use
in supercritical regimes. 
       \begin{figure}
      \includegraphics[width=0.5\textwidth]{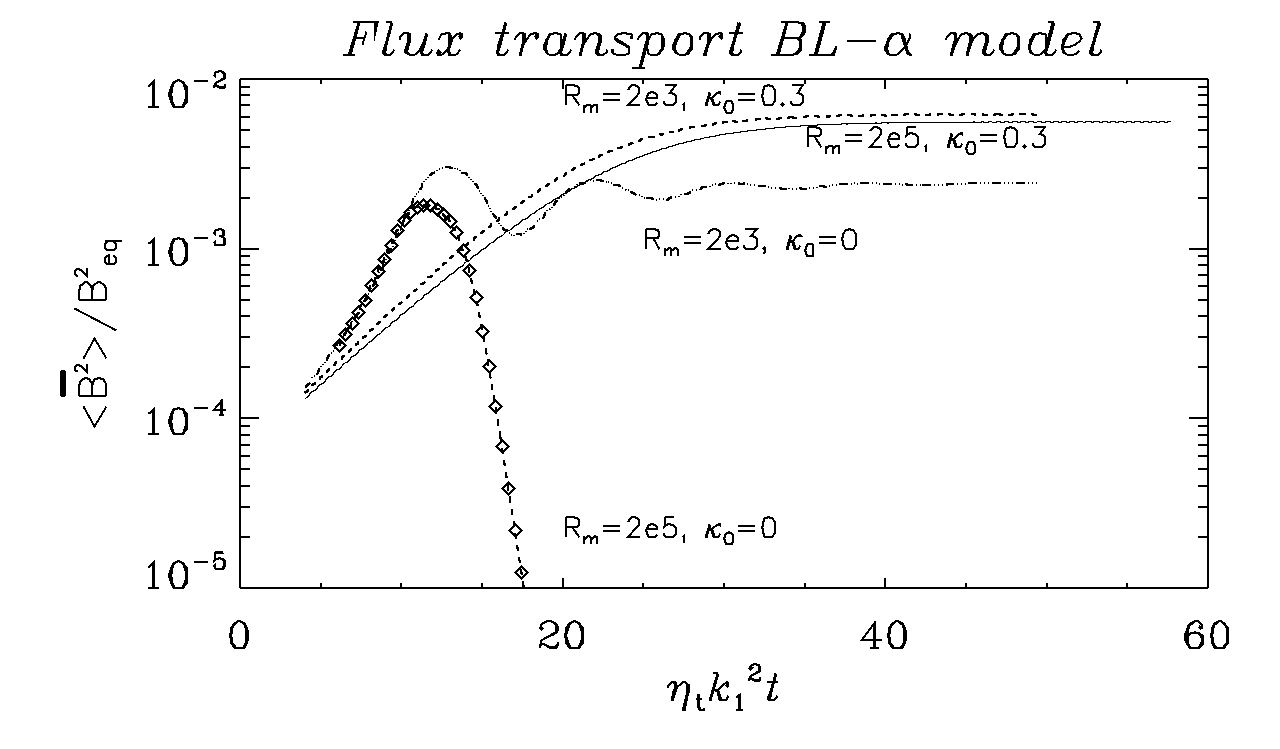}
      \caption{$B^2/B_{\rm eq}^2$ for 
      the flux transport dynamo model of \S3.2 for $R_{\rm m} = 2\times10^3$ with $\kappa_0=0.3$ (dashed); $R_{\rm m} = 2\times10^5$ with $\kappa_0=0.3$ (solid);
      $R_{\rm m} = 2\times10^3$ with $\kappa=0$ (dashed-dotted); $R_{\rm m} = 2\times10^5$ with $\kappa_0=0$ (diamond+dashed).
              }
         \label{fig:energy5}
   \end{figure}
   
   At first we artificially turn off the advective flux 
due to meridional circulation as well as the diffusive flux only in Eq.~(\ref{eq:alphaeq}),
while having them in the induction equations for
$B_{\phi}$ and $A_{\phi}$. The saturation curve for $\Rm=2\times10^3$
is now over-damped whereas the dynamo fails to generate a finite $B_{sat}$
for $\Rm=2\times10^5$ even though it initially has the same growth rate. 
On increasing $\alpha_{BL}=10$ ms$^{-1}$ from 6 ms$^{-1}$ the 
saturation curve for $\Rm=2\times10^5$ also displays overdamped behaviour. 
This indicates that the total $\alpha$ in the domain was simply becoming
sub-critical and the dynamo was not able to sustain itself through 
the saturation phase. We show the distribution of magnetic helicity
in the meridional plane for in Fig.~\ref{fig:mc1}a, b. Note that $\alpha_M$
inside the domain is larger for $\Rm=2\times10^5$ compared to 
$\Rm=2\times10^3$ for the same value of $\alpha_{\rm BL}$.

Inclusion of meridional circulation in Eq.~(\ref{eq:alphaeq}) means that 
we also require a diffusive flux in Eq.~(\ref{eq:alphaeq}) to keep the 
system numerically stable. 
A diffusive flux in this equation is known to alleviate catastrophic 
quenching in $\alpha^2$ (Brandenburg et al.\ 2009) 
as well as $\alpha\Omega$ dynamos (Guerrero, Chatterjee \& Brandenburg 2010). 
It is clear from Fig.~\ref{fig:energy5} that
the overdamped behaviour after the end of the kinematic phase is
suppressed due to a diffusive flux of $\alpha_{\rm M}$ which essentially
reduces the effective decay time for $\alpha_{\rm M}$ to much less 
than $R_{\rm m}/\eta_{\rm t} k_{\rm f}^2$. It may be noted that the dependence of the saturation value of the magnetic energy on $\Rm$ is now much weaker
than the corresponding variation without fluxes. In presence of 
diffusive and advective fluxes due to meridional circulation in
Eq.~(\ref{eq:alphaeq}) the small-scale helicity is distributed through
out the convection zone as shown in Fig.~\ref{fig:mc2}a, b. It is instructive 
to compare the weaker magnitudes of $\alpha_{\rm M}$ with Fig.~\ref{fig:mc1}. 

The diffusive fluxes are therefore crucial
for operation of a successful mean field $\alpha\Omega$ dynamo. 
However an interesting observation is the distribution of $\alpha_M$ 
in the concentrated region at the lower part
of the convection zone (see Fig.~\ref{fig:mc1}) in contrast to 
Fig.~\ref{fig:mc2}. Even though $\alpha_{BL}$ is a surface phenomena, 
considerable magnetic helicity is generated when the meridional 
circulation sinks the poloidal field lines at 
high latitudes and brings them near the tachocline
where toroidal fields are generated. 
\begin{figure}
\includegraphics[width=0.5\textwidth]{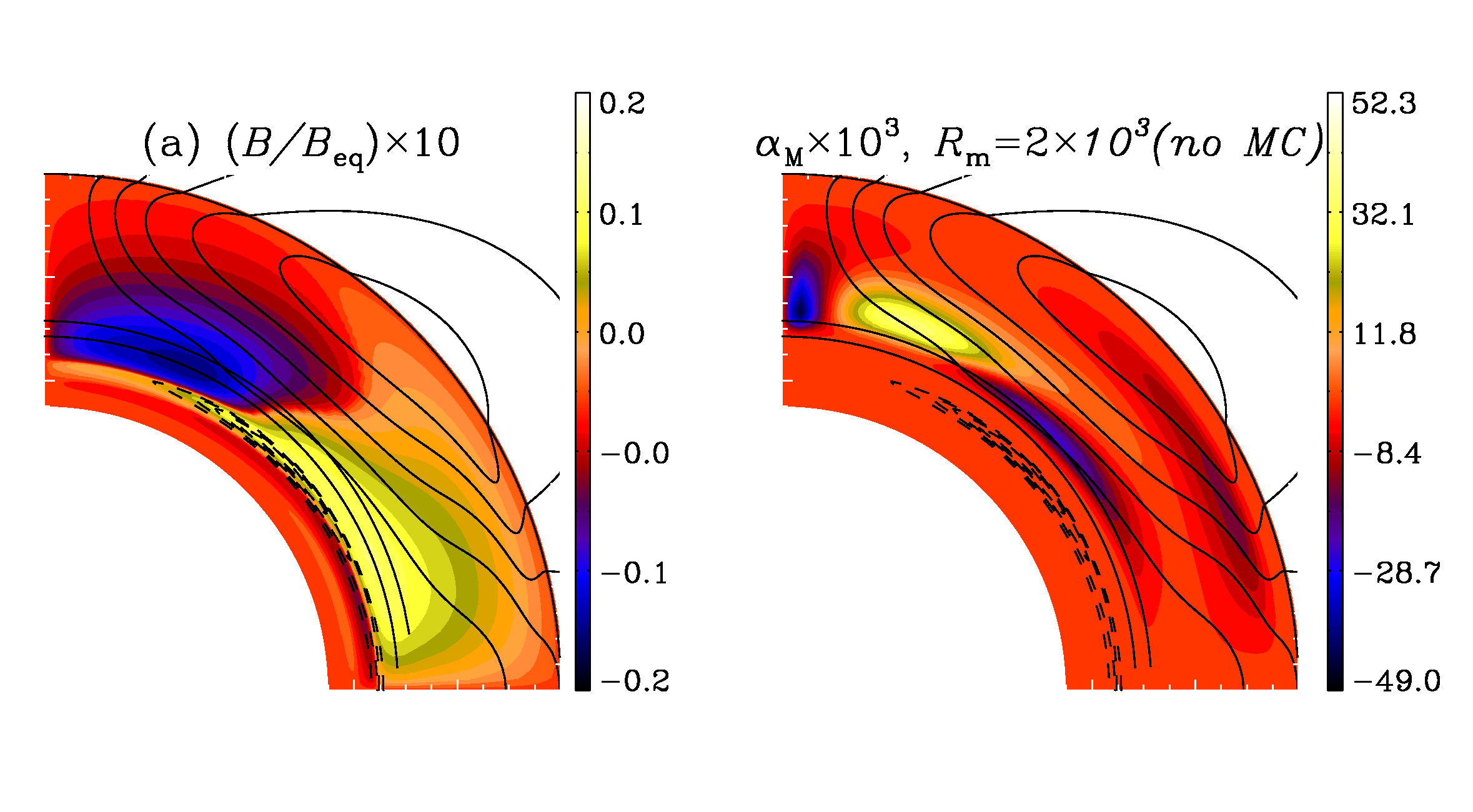}
\includegraphics[width=0.5\textwidth]{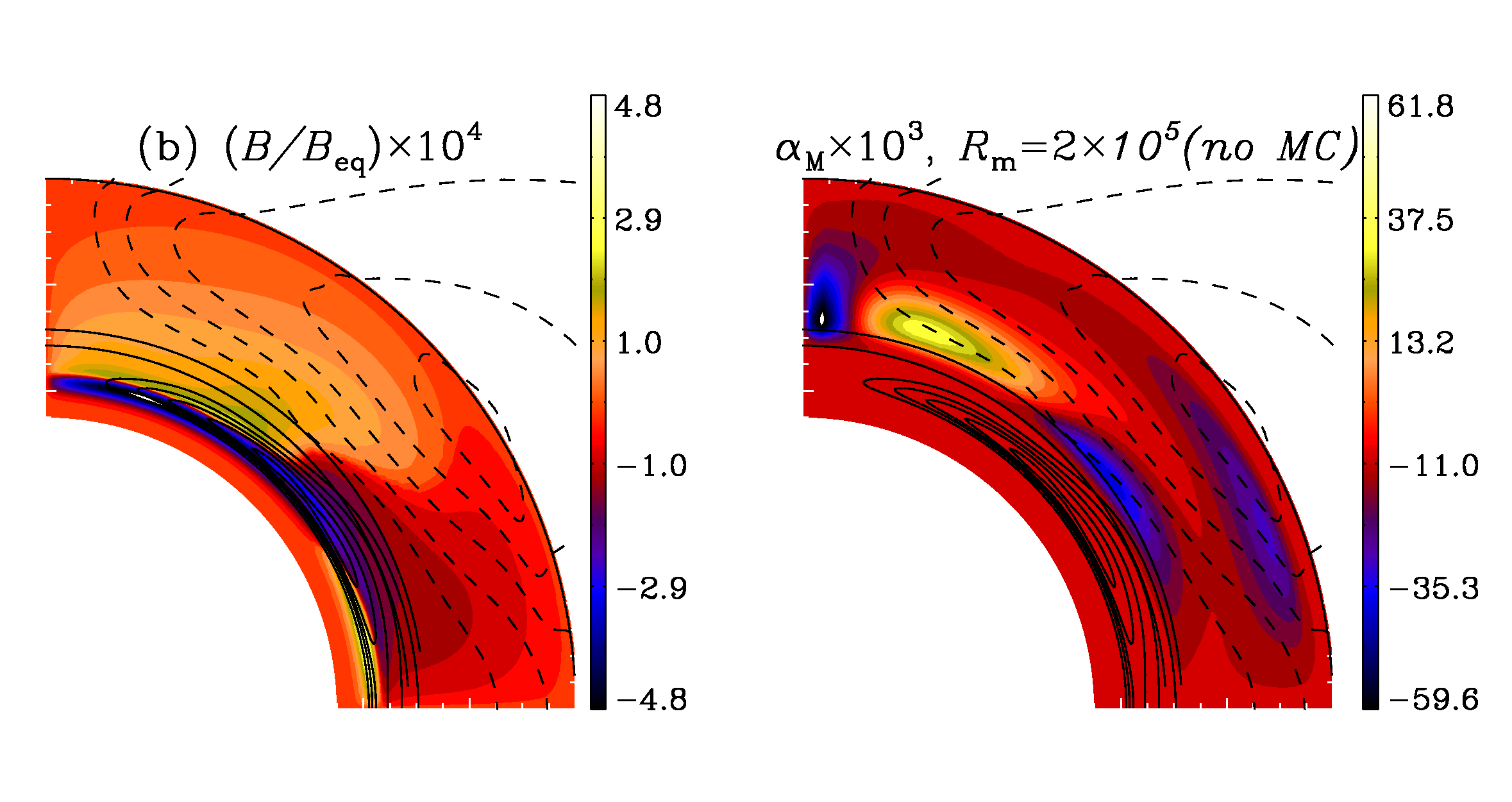}
\label{fig:mc1}
\caption{\label{fig:mc1} Meridional cross-sections showing the distribution 
of toroidal field and $\alpha_M$ for a Babcock-Leighton dynamo 
{\em without} MC and diffusive helicity fluxes in Eq.~\ref{eq:alphaeq} 
for (a) $\Rm=2\times10^3$ and (b) $\Rm=2\times10^5$. The streamlines of the positive and negative 
poloidal field are shown by solid and dashed lines respectively. Note that
the magnetic field has decayed to very small values for $\Rm=2\times 10^5$.}
\end{figure}
\begin{figure}
\includegraphics[width=0.5\textwidth]{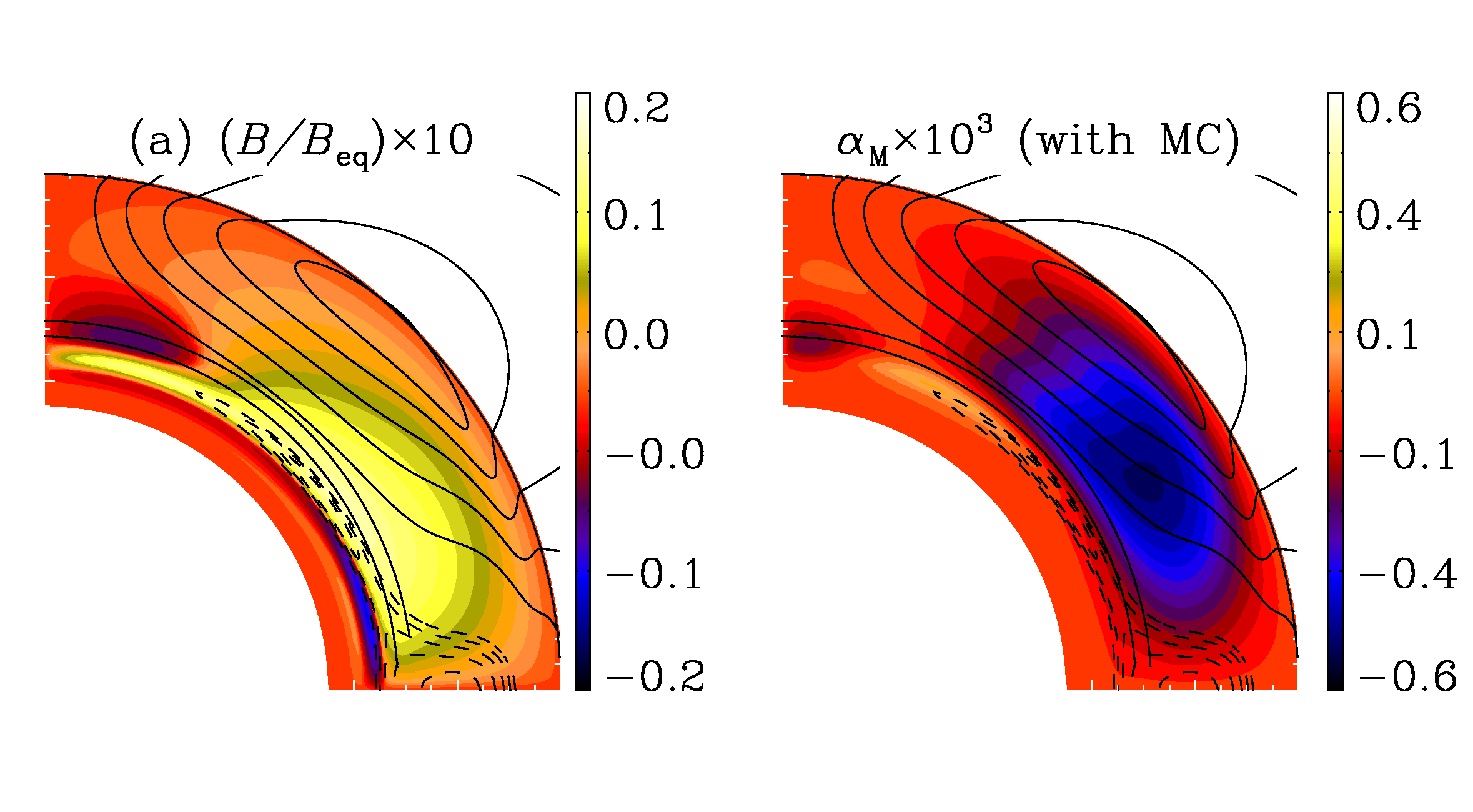}
\includegraphics[width=0.5\textwidth]{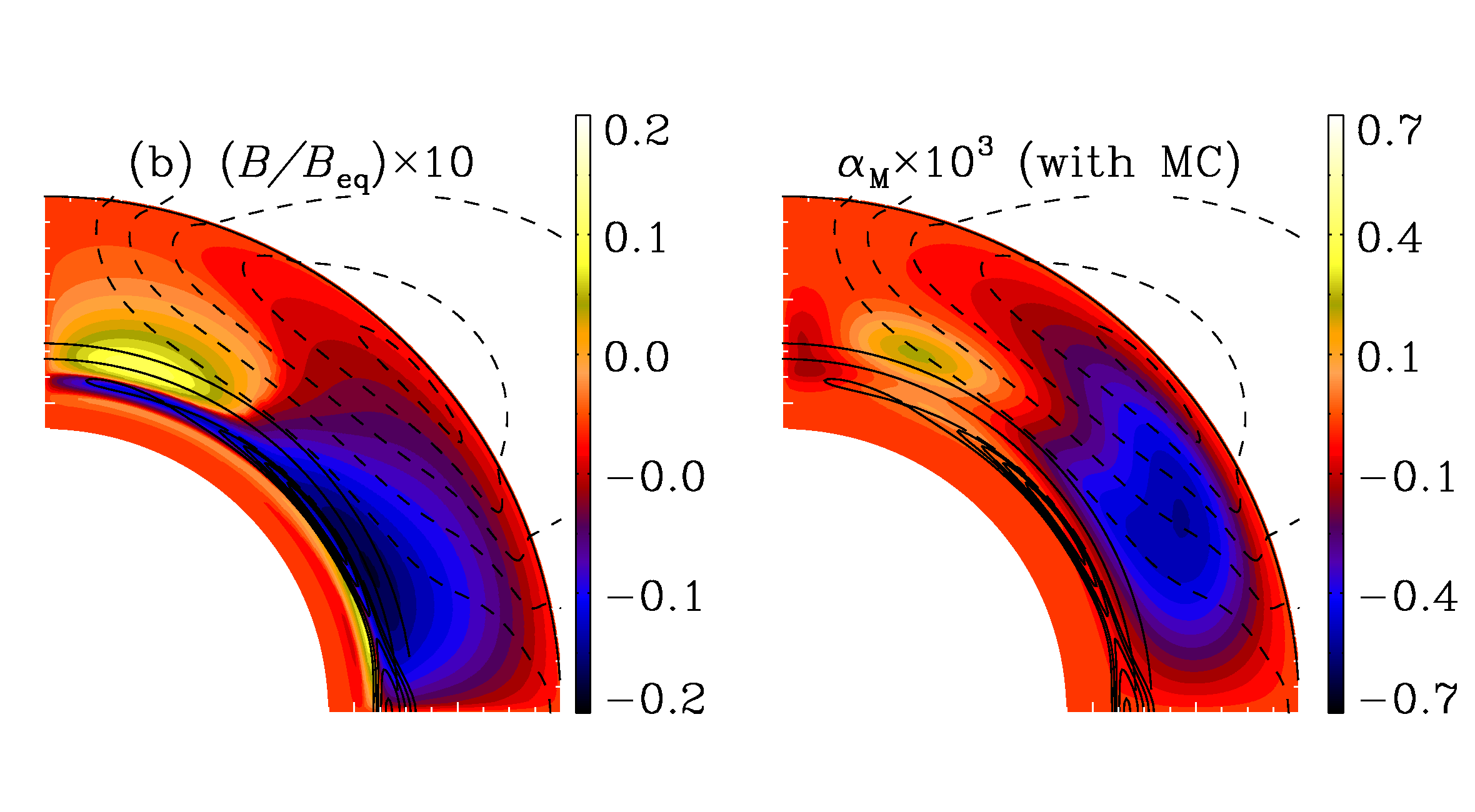}
\label{fig:mc2}
\caption{\label{fig:mc2} Meridional cross-sections showing the distribution 
of toroidal field and $\alpha_M$ for a Babcock-Leighton dynamo 
{\em with} MC and diffusive helicity fluxes for $\Rm=2\times10^3$ at 
two different epochs. The streamlines of the positive and negative 
poloidal field are shown by solid and dashed lines respectively.}
\end{figure} 

%
\section{Conclusions}

We have performed calculations for $\alpha\Omega$ dynamos in a 
spherical shell for spatially segregated $\alpha$ and $\Omega$ 
source regions. The two classes of models we have studied resemble the
Parker's interface dynamo and the Babcock-Leighton dynamo. 

In agreement with earlier work, it is not
possible to escape catastrophic quenching by merely separating the regions of 
shear and $\alpha$-effect.
The saturation value of magnetic energy decreases as $\sim R_{\rm m}^{-1}$ for 
both dynamical quenching and the algebraic quenching with $g_{\alpha}=\Rm$ for 
the simple two layer model without meridional circulation (Fig.~\ref{fig:satrm}). However we find 
that a richer dynamical behaviour emerges for the cases with 
dynamical $\alpha$ effect, in terms of parity fluctuations and 
appearance of `secondary' dynamos (Fig.~8, 9).
We do not see evidence for
chaotic behaviour in the time series of magnetic energy since the dynamo
period and the saturation energy remains fairly constant. However this may not
be the case in presence of diffusive helicity fluxes which introduce further
complexity to the system.
Addition of diffusive helicity fluxes relaxes the catastrophic $\Rm^{-1}$
dependence of the saturation magnetic energy (Fig.~10a, 12).
An interesting `side-effect' of diffusive helicity fluxes is the
appearance of poleward propagating secondary dynamos.
However, because of the lack of scale separation between the mean
field and the forcing scale of the helical turbulence we refrain from
interpreting this in terms of the poleward migration seen in the Sun.
It remains to explore
the role of the solar wind, coronal mass ejections 
which might help in throwing out the small 
scale helicity from the Sun and thus alleviate catastrophic quenching. 
The effects of Vishniac \& Cho fluxes have been investigated and were found to be
of secondary importance compared to diffusive helicity fluxes  
for $\alpha\Omega$ mean field dynamos
(Guerrero, Chatterjee \& Brandenburg 2010).

When both the meridional 
circulation and the diffusive helicity fluxes are artificially 
shut off in the helicity evolution equation, 
the dynamo fails to reach significant saturation values, as expected (Fig.~12).
It is interesting that the Babcock-Leighton dynamos, where $\alpha$ is 
concentrated only in a narrow layer at the surface, also produce 
considerable helicity inside the convection zone when the dynamical quenching (Eq.~\ref{eq:alphaeq}) is employed (Fig.~13, 14).

We have to be cautious about using dynamical quenching equation for 
dynamo numbers not very large compared to the critical dynamo number.
For highly supercritical $\alpha$, the behaviour of the system begins
to be governed by $\alpha_{\rm M}$. We would expect that the magnetic field
should affect all the turbulent coefficients including both
$\alpha$ and $\eta$. However for this analysis we have not included 
an equation for the variation for $\eta_{\rm t}$. This is justified for the simple
two layer model with a lower $\eta_{\rm t}$ in the region of production of strong 
toroidal fields and a higher $\eta_{\rm t}$ in the region of weaker poloidal 
fields. 
It may also be noted that by quenching the diffusivity inversely with the
magnetic energy in a nonlinear dynamo model, Tobias (1996) was able to produce a
bonafide interface model where the magnetic field was restricted to a thin 
layer at an interface between a layer of shear and cyclonic turbulence. However
none of the previous interface models have used the dynamical quenching equation. 

Unfortunately the direct numerical simulations have not yet reached the
modest Reynolds numbers used in this paper ($\sim 10^4$) which are still much lower 
than the astrophysical dynamos. To verify if the equation for dynamical 
quenching works in the same way as in $\alpha^2$ dynamos, we need to embark
upon systematic comparisons between DNS with shear and convection 
and mean field modelling for $\alpha\Omega$ dynamos. 

\begin{acknowledgements}
This work was supported in part by
the European Research Council under the AstroDyn Research Project No.\ 227952
and the Swedish Research Council Grant No.\ 621-2007-4064.
\end{acknowledgements}

\newcommand{\yan}[3]{ #1, {AN,} {#2}, #3}
\newcommand{\yana}[3]{ #1, {A\&A,} {#2}, #3}
\newcommand{\yapj}[3]{ #1, {ApJ,} {#2}, #3}
\newcommand{\yjfm}[3]{ #1, {JFM,} {#2}, #3}
\newcommand{\ymn}[3]{ #1, {MNRAS,} {#2}, #3}
\newcommand{\ysph}[3]{ #1, {Sol. Phys.,} {#2}, #3}
\newcommand{\ypre}[3]{ #1, {PRE,} {#2}, #3}

\end{document}